\documentclass[aps,prb,twocolumn,nofootinbib,longbibliography]{revtex4-2}

\usepackage{amsmath}
\usepackage{graphicx}
\usepackage{braket}
\usepackage{lipsum}
\usepackage{mwe}
\usepackage{caption}
\usepackage{comment}
\newcommand{\ii}{\mathrm{i}}

\newcommand{\brakets}[2]{\left \langle #1 \middle| #2 \right \rangle}
\usepackage[inkscapeformat=png]{svg}
\usepackage{multirow}

\usepackage{booktabs}

\usepackage{microtype} 
\usepackage[]{newtxtext} 
\usepackage[]{newtxmath}

\usepackage{hyperref}
\hypersetup{colorlinks,
	linkcolor={blue!75!black!80!yellow},
	citecolor={blue!75!black!80!yellow},
	urlcolor={blue!75!black!80!yellow}
}
\begin{document}

\title{Higher-order Topological Knots\\ and the classification of non-Hermitian lattices under $C_n$ symmetry}
\author{Yifan Wang}
\author{Wladimir A. Benalcazar}
\email{benalcazar@emory.edu}
\affiliation{Department  of  Physics,  Emory  University, Atlanta, Georgia 30322, USA}
\begin{abstract}
    In two dimensions, Hermitian lattices with non-zero Chern numbers and non-Hermitian lattices with a higher-order skin effect (HOSE) bypass the constraints of the Nielsen–Ninomiya ``no-go'' theorem at their one-dimensional boundaries. This allows the realization of topologically-protected one-dimensional edges with nonreciprocal dynamics. However, unlike the edge states of Chern insulators, the nonreciprocal edges of HOSE phases exist only at certain edges of the two-dimensional lattice, not all, leading to corner-localized states. In this paper, we investigate the topological connections between these two systems and uncover novel non-Hermitian topological phases possessing ``higher-order topological knots'' (HOTKs). These phases arise from multiband topology protected by crystalline symmetries and host nonreciprocal edge states that circulate the entire boundary of the two-dimensional lattice. We show that phase transitions typically separate HOTK phases from ``complex Chern insulator'' phases---non-Hermitian lattices with nonzero Chern numbers protected by imaginary line gaps in the presence of time-reversal symmetry.
\end{abstract}

\maketitle

\tableofcontents
\section{Introduction}
The discrete translation invariance of crystals yields profound implications for the fields they support.
One such implication is articulated by the Nielsen and Ninomiya (NN) theorem, which asserts the impossibility of constructing Hermitian lattice models for non-self-interacting chiral fermions~\cite{NIELSEN198120,NIELSEN1981173,cmp/1103921543,NIELSEN1981219}. In one-dimensional (1D) lattices, this theorem implies that a linear Hermitian system cannot exhibit an imbalance between the number of right-moving and left-moving particles.

Chern insulators circumvent the NN theorem at the boundaries of two-dimensional (2D) lattices, where edge states exhibit unidirectional propagation. Similarly, 2D lattices with a higher-order skin effect (HOSE) display edges with unidirectional propagation, albeit not along all edges \cite{kawabata_higher-order_2020,okugawa_second-order_2020,liu_second-order_2019,lee_hybrid_2019}.

In both cases, a bulk-boundary correspondence connects the nontrivial topological properties of bulk states to unconventional spectral characteristics of edge states across the Brillouin zone.

\begin{figure}
    \centering
    \includegraphics[width=\columnwidth]{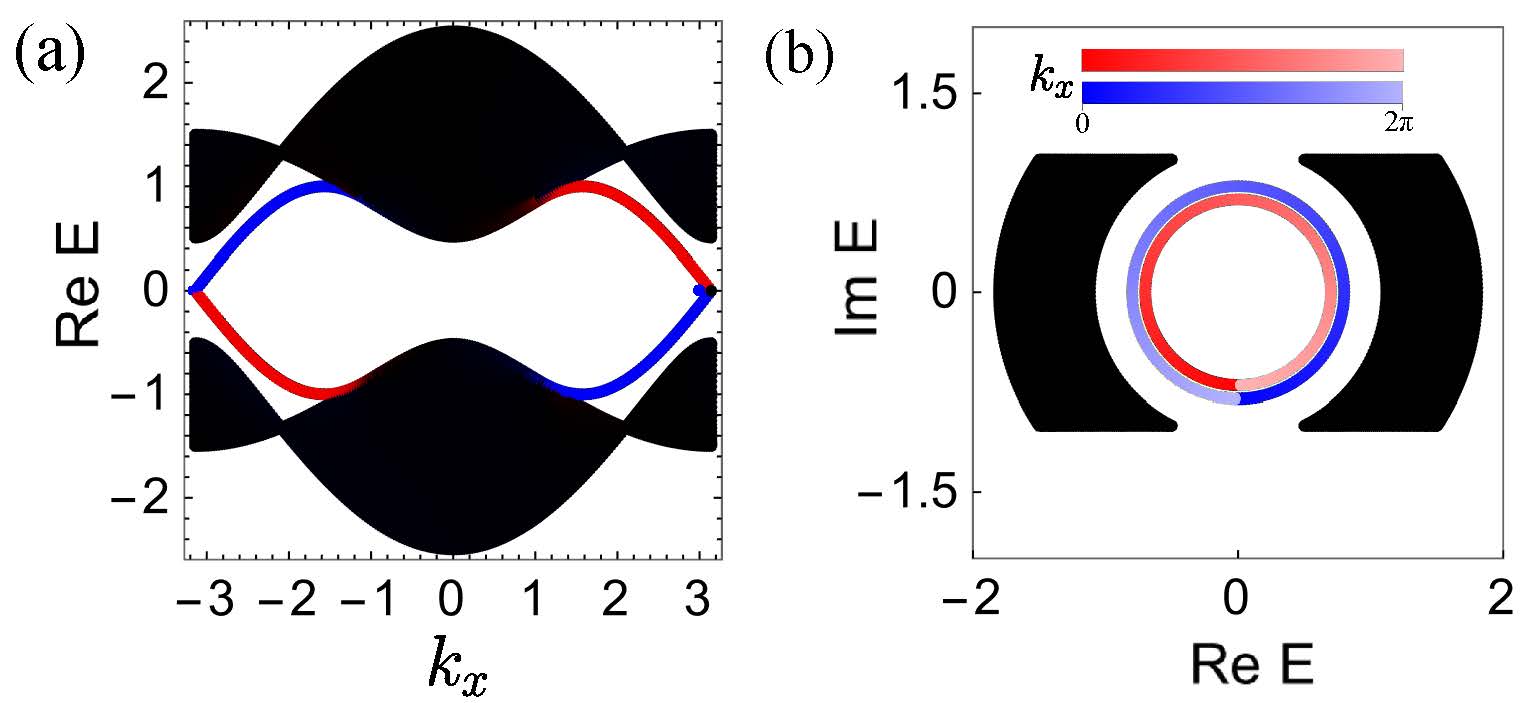}
    \caption{
    Energy spectra under PBC along $x$ and OBC along $y$ of (a) a Chern insulator and (b) a HOSE phase. Bulk bands are depicted in black, while red and blue represent states localized at opposite boundaries. In (b), the brightness of the red and blue colors indicates the value of the crystal momentum $k_x$. As indicated in the color bar, the colors are brightest at $k_x=0$, and gradually fade as $k_x$ approaches $2\pi$. The red and blue spectra in (b) are degenerate; they have been slightly offset for clarity.}
    \label{fig:Intro}
\end{figure}

For Chern insulators, the bulk invariant is the Chern number, which ensures that edge states connect with bulk states in certain regions of the spectrum [Fig.~\ref{fig:Intro}(a)]. For HOSE phases, the precise bulk invariant has not been completely elucidated~ \cite{kawabata_higher-order_2020,Shiozaki_2021}. Furthermore, the edge states in HOSE phases are completely separated from the bulk states by a gap, and their topological nature can be determined by the winding number of the edge complex spectrum across the Brillouin zone alone [Fig.~\ref{fig:Intro}(b)].

The nonreciprocal nature of the edge states of Chern insulators and (some of) the edges of HOSE phases motivates us to look into their topological connections. Specifically, what is the minimal sequence of phase transitions that separate these two phases, and how do the edge states of a Chern insulator spectrally separate from its bulk bands as it transitions into a HOSE phase? Moreover, can this process provide insights into the bulk-boundary correspondence in HOSE phases?

In this paper, we address these questions. We begin by demonstrating how the minimal model of a Chern insulator deforms into a HOSE phase. During this deformation, the edge states of the Chern insulator detach from the bulk, evolving into the non-Hermitian (NH) edge states with nontrivial spectral winding characteristic of the HOSE phase. However, this detachment occurs only along one pair of opposite edges; at the other pair, the states merge into the bulk at the transition. Under full OBC, the nontrivial winding of the detached edge states manifests as a skin effect, collapsing these states into the $O(L)$ corner-localized states that characterize this phase. 

Next, we present NH topological phases, which we term ``higher-order topological knot'' (HOTK) phases. These phases are characterized by bulk states with a line gap and edge states with point gaps and nontrivial windings \emph{along all edges}. Unlike the skin effect of HOSE phases, the edge states in HOTK phases under OBC circulate along the entire perimeter of the sample. We show that the minimal HOTK phase can be generated from lattices with ``complex Chern bands" via topological phase transitions that close a real line gap.

Throughout our analysis, we mostly consider NH Hamiltonians. The topological classification of NH Hamiltonians is largely expanded from the 10-fold classification of Hermitian Hamiltonians~\cite{altland_nonstandard_1997,chiu_classification_2016,kitaev_periodic_2009,Teo_2010,Ryu_Schnyder_Furusaki_Ludwig_2010b,Schnyder2009,Stone_Chiu_Roy_2010}. This expansion arises because the three basic symmetries --- time-reversal ($\mathcal{T}$), particle-hole ($\mathcal{P}$), and chiral ($\mathcal{C}$) --- split into two distinct versions. For instance, time-reversal symmetry (TRS) bifurcates into the conventional TRS, under which a Bloch Hamiltonian $h({\bf k})$ satisfies
\begin{equation}
    \mathcal{T}h(\textbf{k})\mathcal{T}^{-1}=h(-\textbf{k}),
    \label{eq:regular TRS}
\end{equation}
and an additional ``pseudo-TRS'', denoted here as TRS$^\dagger$,
\begin{equation}
    \mathcal{T}h(\textbf{k})\mathcal{T}^{-1}=h(-\textbf{k})^\dag,
    \label{eq:pseudo TRS}
\end{equation}
where ${\bf k}$ is the crystal momentum, $\mathcal{T}=U\mathcal{K}$ is the time reversal operator, $U$ is a unitary matrix and $\mathcal{K}$ denotes complex conjugation.
Taking into account the ramification of TRS, particle-hole symmetry, and chiral symmetry, as well as the additional property of pseudo-Hermiticity, $\eta h({\bf k}) \eta^{-1}=h({\bf k})^\dagger$, where $\eta$ is unitary, the 10-fold classification of Hermitian Hamiltonians expands into a 38-fold classification for NH Hamiltonians~\cite{kawabata_symmetry_2019,ZZhou_2019,Bernard_2002}.

In the absence of pseudo-Hermiticity, a NH Hamiltonian necessarily breaks either TRS or TRS$^\dagger$. Systems exhibiting a skin effect can satisfy~\eqref{eq:regular TRS}, but they must break pseudo-TRS~\eqref{eq:pseudo TRS} (Appendix \ref{Appendix:winding number and TRS}). Here, we focus on 2D lattices belonging to class AI in the 38-fold classification. Class AI encompasses Hamiltonians that obey TRS \eqref{eq:regular TRS} with $\mathcal{T}^2=1$.

The topological classification for class AI is indicated in Table~\ref{tab:TopologicalClassification}. In 1D, NH Hamiltonians with a point gap are topologically classified by a $\mathbb{Z}$ invariant, the winding number. Systems with a real line gap are always trivial, whereas those with an imaginary line gap possess a $\mathbb{Z}_2$ classification, where the topological invariant is the Berry phase, constrained to take values of $0$ or $\pi$ (Appendix~\ref{Appendix:Z2 invariant}). In 2D, there are nontrivial classes only in the presence of an imaginary line gap. The corresponding invariant is the Chern number, which protects ``complex Chern bands'', bands in the complex energy plane that possess nonzero Chern numbers and support edge states that traverse either a real line gap [Fig.~\ref{fig:ComplexChernInsulators}(a)] or an imaginary line gap [Fig.~\ref{fig:ComplexChernInsulators}(b)].

\begin{table}
    \centering
    \begin{tabular}{lcc}
        \toprule
         Gap & 1D & 2D\\
         \midrule
         Point gap&  $\mathbb{Z}$& 0\\
         Real line & $0$ & $0$\\
         Imaginary line & $\mathbb{Z}_2$ & $\mathbb{Z}$\\
         \bottomrule
    \end{tabular}
    \caption{Topological classification in class AI of the 38-fold classification of NH Hamiltonians in 1D and 2D lattices.}
    \label{tab:TopologicalClassification}
\end{table}

The classification presented in Table~\ref{tab:TopologicalClassification} pertains only to first-order topological phases. In the presence of additional crystalline symmetries, the classification is further expanded to account for the protection of higher-order topological phases. This protection is weaker, however, as crystalline symmetries are not local, and disorder in the lattice can disrupt them. Nonetheless, the topological properties of the disordered system persist as long as the energy gap remains open.

We consider NH Hamiltonians obeying $C_n$ symmetry,
\begin{equation}
    \hat{r}_n h(\textbf{k})\hat{r}_n^{-1}=h( R_n\textbf{k}),
\end{equation}
where $\hat{r}_n$ is the rotation operator satisfying $\hat{r}_n^n=1$ (or $\hat{r}_n^n=-1$ in the presence of magnetic fields threading the lattice), and $R_n$ is the matrix that rotates the crystal momentum $\textbf{k}$ by $2\pi/n$ rad.

We will see that HOTK phases exhibit bands in the complex energy plane with line gaps, yet they undergo phase transitions characterized by point gaps. These ``critical point gaps'' bifurcate the topological edge states with point gaps and nontrivial winding as the system transitions from a trivial to a HOTK phase. In the bulk, the HOTK topology is diagnosed by symmetry indicator invariants under $C_n$ symmetry, for which we build the complete classification. The analysis of the symmetry indicator invariants reveals that HOTK phases result from multiband topology, which is possible in NH systems due to the simultaneous presence of multiple line gaps in the complex energy plane.

The structure of the paper is as follows. In Sec.~\ref{Sec: complex Chern insulators}, we describe complex Chern insulators in the presence of TRS. In Sec.~\ref{sec:HOSE}, we overview the higher-order skin effect. In Sec.~\ref{sec:DeformationComplexChern}, we draw connections between Chern insulators and HOSE phases to motivate our study. Section~\ref{Sec: HOTK edge current} presents our main findings; we classify NH Hamiltonians according to $C_n$ symmetry, and introduce models with HOTK phases, one for each of $C_{2,3,4,6}$ symmetries. Finally, in Sec.~\ref{sec:discussion and conclusion} we conclude with a discussion and outlook of our paper.

\section{Complex Chern insulators \\ under time-reversal symmetry}\label{Sec: complex Chern insulators}
\begin{figure}
    \centering
    \includegraphics[scale=0.8]{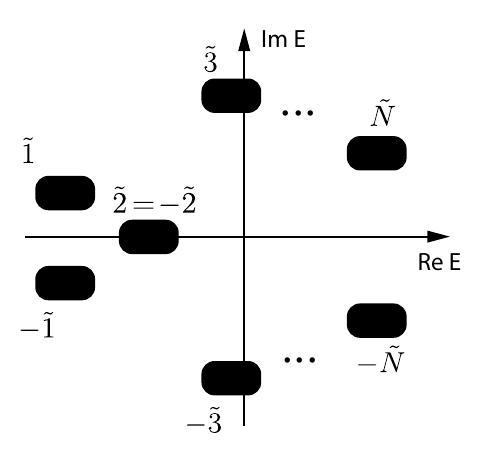}
    \caption{Schematic of the energy spectrum of a generic NH Hamiltonian in class AI. Energy bands that map into one another by time reversal are labeled in pairs $-\tilde{n}, \tilde{n}$.}
    \label{fig:Labeling Bands}
\end{figure}

The Chern number of a Hermitian system vanishes under TRS~\eqref{eq:regular TRS}. However, this is not always the case for NH systems. 
Consider a NH Bloch Hamiltonian $h({\bf k})$. The right and left eigenstates obey
\begin{align}
&h(\textbf{k})\ket{u_\textbf{k}^n}=\epsilon_n(\textbf{k})\ket{u_\textbf{k}^n},\nonumber\\
&h^\dag(\textbf{k})\ket{v_\textbf{k}^n}=\epsilon^*_n(\textbf{k})\ket{v_\textbf{k}^n},
\end{align}
respectively. They can be made to obey $\brakets{u^m_{\bf k}}{v^n_{\bf k}}=\delta_{mn}$ and $\sum \ket{u^m_{\bf k}}\bra{v^m_{\bf k}}=1$.
A NH Hamiltonian in class AI obeys \eqref{eq:regular TRS}. Applying $h(-\textbf{k})\mathcal{T}$ on a right eigenstate of $h(\textbf{k})$ we have
\begin{equation}
    h(-\textbf{k})\mathcal{T} \ket{u_{\textbf{k}}^n}=\mathcal{T}h(\textbf{k})\ket{u_{\textbf{k}}^n}=\epsilon_n^*(\textbf{k})\mathcal{T}\ket{u_{\textbf{k}}^n}.
    \label{eq:Time reversal 1}
\end{equation}
Hence, $\mathcal{T}\ket{u_{\textbf{k}}^n}$ is an eigenstate of $h(-\textbf{k})$ with the eigenvalue $\epsilon_n^*({\bf k})$ and, as such, is proportional to $\ket{u^n_{- {\bf k}}}$, which has energy $\epsilon_n(-{\bf k})$. Thus, under TRS \eqref{eq:regular TRS}, the eigenvalues come in pairs $\{\epsilon_n(-\textbf{k}),\epsilon^*_n(\textbf{k})\}$. 
This allows us to label the energy bands using the following notation: let $\Tilde{1},\Tilde{2},\cdots, \Tilde{N}$ denote energy bands above $\text{Im} E=0$ and $-\Tilde{1},-\Tilde{2},\cdots, -\Tilde{N}$ denote the corresponding bands below $\text{Im} E=0$, such that bands $\Tilde{n}$ and $-\Tilde{n}$ are related by TRS. If a band $\Tilde{l}$ lies on the real energy line, we say that $\Tilde{l}=-\Tilde{l}$. This notation is schematically represented in Fig.~\ref{fig:Labeling Bands}. 
 
At a given $\textbf{k}$, let the pair of indices $n\in \Tilde{n}$ and $-n\in -\Tilde{n}$ denote two states on opposite sides of the imaginary line gap. Choosing a gauge in which the sewing matrix $V_{\textbf{k}}^{-n,n}=\bra{v_{-\textbf{k}}^{-n}}\mathcal{T} \ket{u_\textbf{k}^n}$ is diagonal, we have $\mathcal{T}\ket{u_\textbf{k}^n}=\ket{u_{-\textbf{k}}^{-n}}$; that is, $\mathcal{T}$ takes an eigenstate at ${\bf k}$ and relates it to its time-reversal partner eigenstate at $-{\bf k}$ across the imaginary gap.

Now consider the biorthogonal Berry curvature~\cite{shen_topological_2018} for band~$\Tilde{n}$,
\begin{equation}
    [\Omega_{\Tilde{n}}^{\text{RL}}(\textbf{k})]^{n_1,n_2}=\textrm{i}\left( \brakets{\partial_{k_x}u_\textbf{k}^{n_1}}{\partial_{k_y}v_\textbf{k}^{n_2}}-\brakets{\partial_{k_y}u_\textbf{k}^{n_1}}{\partial_{k_x}v_\textbf{k}^{n_2}}\right),
\end{equation}
where the superscript $\text{RL}$ labels the order of the biorthogonal basis. States $n_1,n_2\in \Tilde{n}$.
Then
\begin{align}
    &[\Omega_{\Tilde{n}}^{\text{RL}}(-\textbf{k})]^{n_1,n_2}=\textrm{i}\left(\brakets{\partial_{k_x}u_{-\textbf{k}}^{n_1}}{\partial_{k_y}v_{-\textbf{k}}^{n_2}}-\brakets{\partial_{k_y}u_{-\textbf{k}}^{n_1}}{\partial_{k_x}v_{-\textbf{k}}^{n_2}}\right)\nonumber \\
    &=\textrm{i}\left(\brakets{\partial_{k_y}v_{-\textbf{k}}^{n_{2}*}}{\partial_{k_x}u_{-\textbf{k}}^{n_{1}*}}-\brakets{\partial_{k_x}v_{-\textbf{k}}^{n_{2}*}}{\partial_{k_y}u_{-\textbf{k}}^{n_{1}*}}\right)\nonumber\\
    &=\textrm{i}\left(\brakets{\partial_{k_y}\mathcal{T}v_{-\textbf{k}}^{n_2}}{\partial_{k_x}\mathcal{T}u_{-\textbf{k}}^{n_1}}-\brakets{\partial_{k_x}\mathcal{T}v_{-\textbf{k}}^{n_2}}{\partial_{k_y}\mathcal{T}u_{-\textbf{k}}^{n_1}}\right)\nonumber\\
    &=\textrm{i}\left(\brakets{\partial_{k_y}v_{\textbf{k}}^{-n_2}}{\partial_{k_x}u_{\textbf{k}}^{-n_1}}-\brakets{\partial_{k_x}v_{\textbf{k}}^{-n_2}}{\partial_{k_y}u_{\textbf{k}}^{-n_1}}\right)\nonumber\\
    &=-[\Omega_{-\Tilde{n}}^{\text{LR}}(\textbf{k})]^{-n_2,-n_1},
\end{align}
where in the first step we used $\brakets{\partial_{k_x}u_{-\textbf{k}}^{n_1}}{\partial_{k_y}v_{-\textbf{k}}^{n_2}}=\brakets{\partial_{k_y}v_{-\textbf{k}}^{n_{2}*}}{\partial_{k_x}u_{-\textbf{k}}^{n_{1}*}}$ and similarly for the second term. We see that $-\Omega_{-\Tilde{n}}^\text{LR}(-\textbf{k})=\Omega_{\Tilde{n}}^{\text{RL}}(\textbf{k})$. These two versions of Berry connections, $\Omega^{\text{LR}}$ and $\Omega^{\text{RL}}$, result in the same Chern number \cite{shen_topological_2018}. Accordingly, the Chern number for bands across an imaginary line gap related to one another by TRS obey
\begin{align}
     C_{\Tilde{n}}&=\frac{1}{2\pi}\int \text{Tr}\left[ \Omega_{\Tilde{n}}^{\text{RL}}(\textbf{k}) \right] d^2\textbf{k} \nonumber\\
    &=-\frac{1}{2\pi}\int \text{Tr} \left[ \Omega_{-\Tilde{n}}^{\text{LR}}(-\textbf{k}) \right] d^2\textbf{k}=-C_{-\Tilde{n}}.
\end{align}
\begin{figure}
    \centering
    \includegraphics[width=\columnwidth]{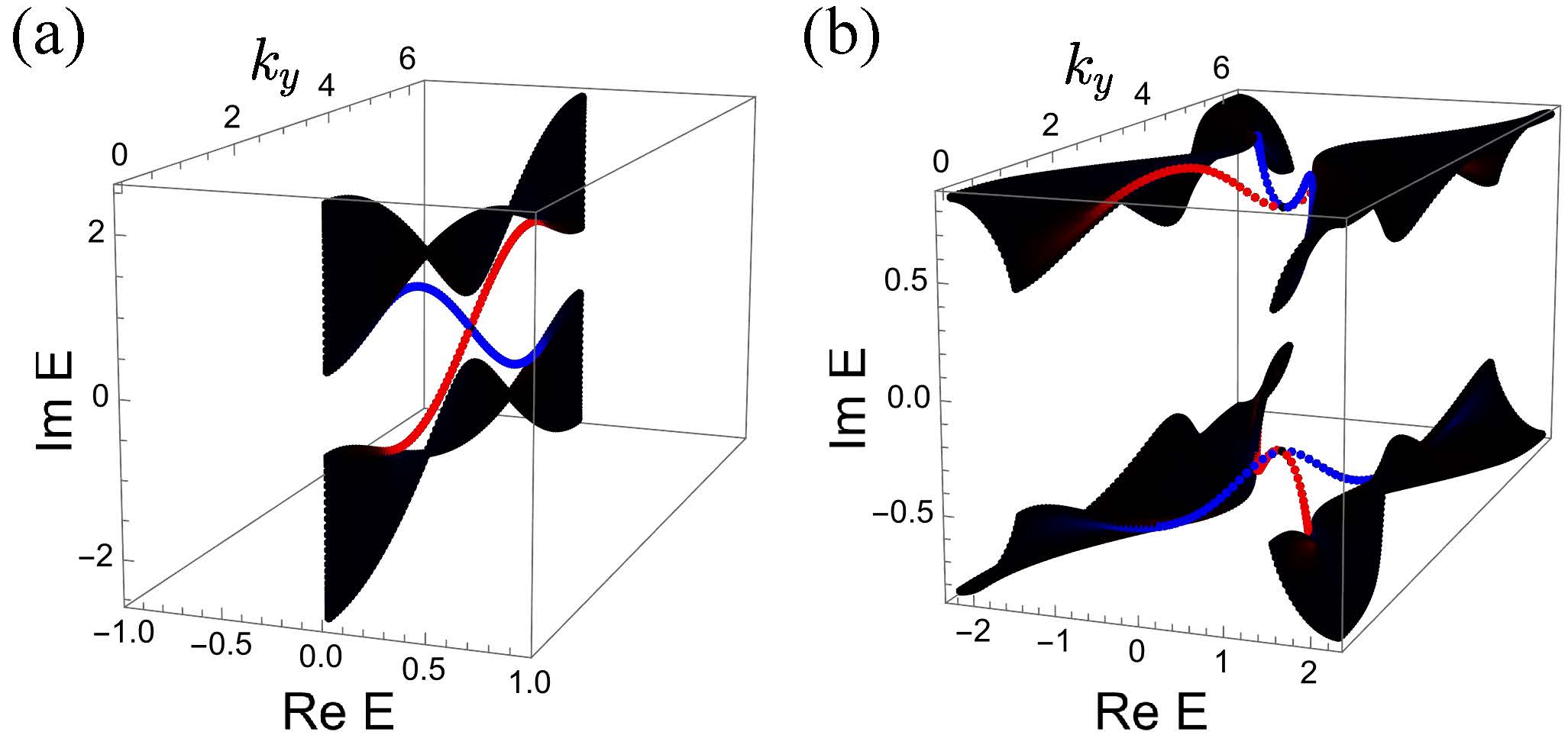}
    \caption{Energies of time-reversal symmetric NH Hamiltonians with ``complex Chern bands'' in the presence of an imaginary line gap. Both spectra are calculated under OBC along $x$ and PBC along $y$. Black bands represent bulk bands, while blue and red lines denote edge states at opposite edges. In (a), the edge states cross an imaginary line gap for Hamiltonian~\eqref{eq:ComplexChernInsulatorImGap} with $m=0.5$, and in (b), they cross a real line gap for Hamiltonian~\eqref{eq:ComplexChernInsulatorReGap} with $t=1.2$. Both systems belong to class AI. Time-reversal symmetry relates the energy bands across the imaginary gap.}
    \label{fig:ComplexChernInsulators}
\end{figure}
The Chern number must only vanish for energy bands lying on the real energy axis, such that $\tilde{n}=-\tilde{n}$. For a minimal model with two bands, $\Tilde{1}$ and $-\Tilde{1}$, related by TRS~\eqref{eq:regular TRS}, we have $C_{\Tilde{1}}=-C_{-\Tilde{1}}$. When such a system is in contact with the vacuum, edge states must close the imaginary line gap that separates bands $\Tilde{1}$ and $-\Tilde{1}$  [Fig.~\ref{fig:ComplexChernInsulators}(a)]. A less intuitive example occurs for a NH Hamiltonian with four bands separated by both real and imaginary line gaps, all having nontrivial Chern numbers; in this case, the edge states can close the real line gap instead [Fig.~\ref{fig:ComplexChernInsulators}(b)]. In both cases, the imaginary line gap closes for a Hermitian system, leading only to trivial phases.

As examples of these phases, first consider the Bloch Hamiltonian 
\begin{align}
    h_\mathrm{CC}^\mathrm{Im}({\bf k})=&-\textrm{i}\sigma_z \sin k_x+\textrm{i} \sigma_x \sin k_y \nonumber\\
    &+\textrm{i} \sigma_y (\cos k_y + \cos k_x+m),
    \label{eq:ComplexChernInsulatorImGap}
\end{align}
where ${\bf k}=(k_x,k_y)$ is the crystal momentum, $\sigma_{x,y,z}$ are the Pauli matrices, and $m$ sets the Chern number $C$, with $C=1$ ($C=-1$) for $0<m<2$ ($-2<m<0$) or $C=0$ else. This model obeys TRS~\eqref{eq:regular TRS} with $\mathcal{T}=\mathcal{K}$. A plot of its energy bands is shown in Fig.~\ref{fig:ComplexChernInsulators}(a) for $m=0.5$. 

Next, consider the Bloch Hamiltonian
\begin{align}
    h_\mathrm{CC}^\mathrm{Re}({\bf k})=& (\cos k_x+t/2) \sigma_x \tau_x + (\cos k_y+t/2)\sigma_z \tau_x \nonumber\\
    &-\sin k_x \sigma_y \tau_x-\sin k_y \sigma_0 \tau_y\nonumber\\
    &+\textrm{i}(t/2)(\sigma_x\tau_y-\sigma_z\tau_y), 
    \label{eq:ComplexChernInsulatorReGap}
\end{align}
where $\sigma_0$ is the $2\times2$ identity matrix. This model obeys TRS~\eqref{eq:regular TRS} with $\mathcal{T}=\mathcal{K}$. When $0<t<1$, edge states cross the imaginary line gap, as in Hamiltonian \eqref{eq:ComplexChernInsulatorImGap}. At $t=1$, all edge bands touch. When $1<t<2$, topological edge states close the real line gap instead. For $t>2$, this model enters the trivial phase, where no edge states exist. Figure~\ref{fig:ComplexChernInsulators}(b) shows the energy bands of Hamiltonian~\eqref{eq:ComplexChernInsulatorReGap} for $t=1.2$. Note that there are two copies of Chern insulators with opposite Chern numbers above and below the imaginary line gap. This is the key feature of a real-line-gap Chern insulators under TRS~\eqref{eq:regular TRS}. The Hamiltonian \eqref{eq:ComplexChernInsulatorReGap} obeys $C_4$ symmetry. Appendix \ref{Sec:NH Breathing Honeycomb} describes a model on a $C_6$-symmetric hexagonal lattice in class AI with a real-line gap Chern insulator phase.

\section{Higher-order skin effect}\label{sec:HOSE}
In addition to bands separated by line gaps, NH systems also exhibit point gaps. In 1D, systems with a point gap exhibit the NH skin effect (NHSE), by which all eigenstates in a crystal with OBC exponentially localize at one of its edges~\cite{shen_topological_2018,yao_edge_2018,Xiong_2018,Song_2019,okuma_topological_2020,Zhang_Yang_Fang_2022,Zhang_2020,Yang_2020,Yokomizi_2019,jiang_2024,Robert-Jan2}.

This effect is associated with a bulk topological invariant, the winding number $W \in \mathbb{Z}$ of the complex energy spectrum with respect to a constant reference point $E_p$ inside the point gap. For a translation-invariant system with PBC, the winding number is defined by 
\begin{align}
    W&=\frac{1}{2\pi\textrm{i}}\int_{BZ} \text{Tr}\left(H'(k)^{-1}d H'(k)\right)\nonumber\\
    &=\frac{1}{2\pi\textrm{i}}\int_{BZ} dk\frac{d}{dk}\log \det H'(k),
    \label{eq:Winding number maintext}
\end{align}
where $H'(k) = H(k)-E_p$ and $E_p$ is any energy inside the point gap. Since the Brillouin zone (BZ) is periodic, the complex spectra of NH Hamiltonians with point gaps form knots (or loops) in $(\mathrm{Re}E,\mathrm{Im}E,k)$ space. When more than one energy band wind, the winding is associated with the braid group $B_N$, where $N$ is the number of separable energy bands~\cite{konig_braid-protected_2023,rui_Hermitian_2023,li_homotopical_2021,wojcik_homotopy_2020,yang_homotopy_2023,hu_knot_2022,hu_knots_2021}. The braid group classifies topologically inequivalent knots or loops. Such classification of 1D NH crystals was developed in Ref.~\cite{hu_knots_2021}. In this and the next section, we focus on the windings and braiding configurations at the 1D boundary of 2D NH lattices for systems with PBC along one direction and OBC along the other, so that, for example, edge states localized at edges $x=1$ or $x=L$ can still be parametrized by the crystal momentum $k_y$ along $y$. As shown in Fig.~\ref{fig:Intro}(b), there are NH lattices in which edge states spectrally disconnect from the 2D bulk bands. Remarkably, while the bulk bands present a line gap, the disconnected edge states present a point gap and carry nontrivial windings [Fig.~\ref{fig:Intro}(b)]. Now, consider the case in which edge windings occur for PBC along $x$ and OBC along $y$, but not vice versa. In that case, under full OBC (i.e., OBC along both $x$ and $y$), only one pair of edges manifests a 1D skin effect, collapsing its edge states to a pair of opposite corners of the 2D crystal. Since for a crystal of $L \times L$ unit cells, $O(L^2)$ states remain distributed across the bulk, and only $O(L)$ states localize at a corner, such phases have been referred to as possessing a ``higher-order skin effect''~\cite{kawabata_higher-order_2020,lee_hybrid_2019,liu_second-order_2019,okugawa_second-order_2020}, in analogy with the existence of $O(1)$ corner states in 2D second-order topological phases~\cite{Benalcazar_2017,Benalcazar_Bernevig_Hughes_science_2017, langbehn2017, song2017,Xue_2021,Hu_2024,Fraxanet_2023}. 

A minimal model of a HOSE phase is given by the Bloch Hamiltonian
\begin{align}
    h_\textrm{HOSE}({\bf k})=&-\textrm{i}(\gamma+\cos k_x)\sigma_0+\sin k_x \sigma_z \nonumber\\
    &+(\gamma+\cos k_y)\sigma_y+\sin k_y \sigma_x,
    \label{eq:HOSE}
\end{align}
proposed in Ref.~\cite{kawabata_higher-order_2020}. The Hamiltonian \eqref{eq:HOSE} can be obtained from the quadrupole topological insulator (QTI)~\cite{Benalcazar_2017}, which is chiral symmetric and can be written as $H_\textrm{QTI}=((0,h_\textrm{HOSE}),(h^\dag_\textrm{HOSE},0))$~\cite{footnote1}. Physically, the Bloch Hamiltonian~\eqref{eq:HOSE} represents 1D horizontal Hatano-Nelson chains with alternating winding numbers stacked along the vertical direction. The chains are coupled via alternating vertical hopping terms with amplitudes 1 and $\gamma$. The phases of Hamiltonian \eqref{eq:HOSE} are controlled by the single parameter $\gamma$. When $0<\gamma<1$, there is a HOSE phase with $O(L)$ corner states at the top-left and bottom-right corners of a square lattice. A phase transition occurs at $\gamma=1$. For $\gamma>1$, the Hamiltonian~\eqref{eq:HOSE} enters the trivial phase where the HOSE vanishes.

\section{Deforming Chern insulators\\ into HOSE phases}
\label{sec:DeformationComplexChern}
Both the edge states of Chern insulators and some of the edges in HOSE phases circumvent the no-go theorem by Nielsen and Ninomiya (NN), which forbids the existence of a net chirality in the states of crystals with noninteracting Hermitian Hamiltonians. The NN theorem stems from a topological consideration; in 1D, it is impossible to make a real spectrum periodic in the BZ without having an equal number of right- and left-moving states. 

Chern insulators circumvent the limitations of the NN theorem by a topological bulk-boundary connection, by which edge states must merge into the bulk for part of the spectrum, as in the region around $k_x=0$ in Fig.~\ref{fig:Intro}(a). There is thus a topological obstruction to spectrally ``peeling off'' the chiral edge states of a Hermitian Chern insulator. At most, a bulk phase transition will eliminate the chiral states altogether. 

\begin{figure}[t!]
    \centering
    \includegraphics[width=\columnwidth]{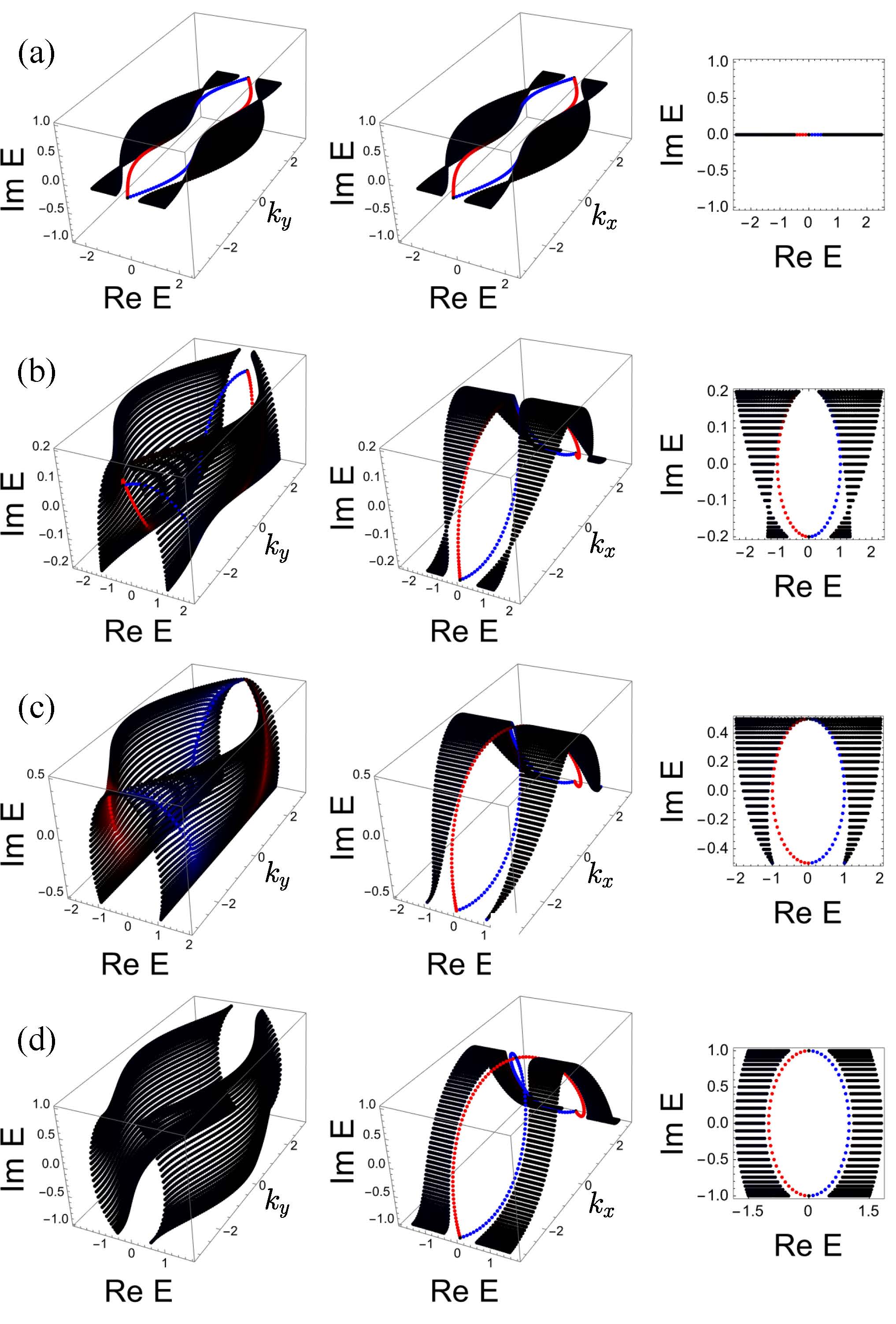}
    \caption{Deforming a real-line-gap Chern insulator into a HOSE phase in the tight-binding model with Bloch Hamiltonian~\eqref{eq:DeformationReGap}. The deformation path chosen in parameter space is $(g,t)=(1,1)-\theta (1,1)$, for $\theta:0 \rightarrow 1$. Panels (a)--(d) correspond to $\theta=0,0.2,0.5,1$, respectively. (a) A Hermitian Chern insulator phase. (b) A real-line-gap complex Chern insulator phase. (c) Bulk phase transition between the Chern insulator phase and the HOSE phase. (d) HOSE phase. In all panels, (left column) OBC along $x$, PBC along $y$; (middle column) OBC along $y$, PBC along $x$; (right column) band projections of the plots on the middle column onto the complex energy plane. Black represents bulk states, while blue and red represent states localized at opposite edges.}
    \label{fig:DeformationReGap}
\end{figure}

Phases exhibiting the skin effect circumvent the limitations of the NN theorem by breaking Hermiticity. As a result, their energy spectra are complex, and can be made periodic in the complex energy plane as the crystal momentum traverses the 1D BZ while preserving a chirality. Since this nontrivial topology exists in a 1D manifold, 1D systems with a point gap topology do not need a bulk of a higher dimension to sustain it, nor a bulk-boundary correspondence. In HOSE phases, the edge states with point-gap topology are spectrally separated from the bulk bands. Yet, the existence of these edge states themselves is a manifestation of a (weak) nontrivial topological configuration of the 2D bulk bands.

The inequivalent ways in which the spectra of a periodic system connect across the BZ --- some of which circumvent the NN theorem and lead to nonreciprocity --- correspond to different topological classes, and thus, crystals that implement them correspond to distinct topological phases of matter, separated by bulk phase transitions. In particular, we are concerned with the relation between Chern insulators and HOSE phases, both of which have chiral edge states. Specifically, we ask whether the edge states of a Chern insulator can be peeled off its bulk if we promote its energy spectrum to the complex plane, in a similar way as the 1D edge states of the HOSE phase do. Such a connection would allow us to establish a relationship between Chern insulators and the HOSE, and consider these different systems on equal footing. 

For this purpose, consider the two-band Bloch Hamiltonian
\begin{align}
    h^\mathrm{Re}_\mathrm{def}({\bf k})= &\sin k_x \sigma_x + \sin k_y \sigma_y\nonumber\\
     &+ (m+t\cos k_x+\cos k_y) \sigma_z\nonumber\\
     &+\textrm{i} \cos k_x(1-g) \sigma_0.
     \label{eq:DeformationReGap}
\end{align}
When $g=1$ and $t=1$, this model is the Qi-Wu-Zhang (QWZ) Hamiltonian~\cite{Qi_2006,Asb_th_2016}, a minimal model for a Chern insulator, with Chern number $C=1$ for $0<m<2$ or $C=0$ otherwise. We fix the mass to $m=0.5$. Its spectrum is shown in Fig.~\ref{fig:DeformationReGap}(a). To deform model \eqref{eq:DeformationReGap} from the Chern insulator phase to the HOSE phase, we continuously vary the parameters $g$ and $t$ according to $(g,t) = \theta (1,1)$ for $\theta: 1 \to 0$.

Note that $g \neq 1$ makes \eqref{eq:DeformationReGap} non-Hermitian, promoting its spectrum to the complex plane. At $\theta=0.2$, the model is in a NH Chern insulator phase [Fig.~\ref{fig:DeformationReGap}(b)], smoothly connected to the QWZ model in Fig.~\ref{fig:DeformationReGap}(a). As the deformation continues, a bulk phase transition at $\theta=0.5$ changes the Chern number from $C=1$ to $C=0$ [Fig.~\ref{fig:DeformationReGap}(c)]. However, not all the chiral edge states merge and disappear into the bulk; instead, along one direction [middle panel in Fig.~\ref{fig:DeformationReGap}(c)], the edge states nontrivially reconnect, changing their topology to now wind in the complex plane. This is evident on the other side of the phase transition, as shown in Fig.~\ref{fig:DeformationReGap}(d) for $\theta=1$.
In Appendix~\ref{Appendix:DeformationImGap}, we present complementary plots to those in Fig.~\ref{fig:DeformationReGap}.

During this deformation, the nontrivial topology in the bulk of the Chern insulator phase that gives rise to the chiral edge states is transferred to a nontrivial topological winding of the edge states themselves, leaving the bulk topologically trivial (from the point of view of strong topology). Since this skin effect occurs only at one pair of edges, a system in this phase with full OBC will necessarily manifest $O(L)$ corner states, i.e., it will manifest a HOSE. In fact, at the end of the deformation process, $(g,t)=(0,0)$, the Hamiltonian \eqref{eq:DeformationReGap} is smoothly deformable to Hamiltonian \eqref{eq:HOSE} up to a transformation $\sigma_x\rightarrow \sigma_z, \sigma_y\rightarrow\sigma_x,\sigma_z\rightarrow\sigma_y$.

A similar deformation can connect the complex Chern insulator with an imaginary line gap \eqref{eq:ComplexChernInsulatorImGap} to a HOSE phase described by Hamiltonian~\eqref{eq:HOSE}. In this case, the entire deformation process is within class AI, obeying TRS \eqref{eq:regular TRS}. This process is shown in Appendix~\ref{Appendix:DeformationImGap}.

\section{Higher-order topological knots}\label{Sec: HOTK edge current}
Since not all the edges in HOSE phases exhibit nonreciprocal transport, these phases possess $O(L)$ states exponentially localized at corners where edges with trivial and nontrivial winding intersect. In contrast, Chern insulators exhibit chiral edge states that extend continuously around the sample, enabling nonreciprocal propagation even in the presence of defects or corners~\cite{Zu_2023,Wang_Chong_Joannopoulos_Soljacic_2009}. Thus, even though the edges in Chern insulators and some of the edges in HOSE phases circumvent the NN theorem, there are important differences between them, resulting in distinct densities of states and associated transport phenomena. These differences raise a question: Can NH lattices sustain nontrivial windings along all edges of a sample, as Chern insulators do, rather than only along some edges, as in HOSE phases?
In the following, we explore lattices with these properties. Specifically, we consider NH bulk-boundary correspondence mechanisms that generate edge states with uniform nontrivial winding across all edges. This approach eliminates the $O(L)$ corner-localized states characteristic of HOSE phases, provided the protecting symmetries remain intact. To this effect, we first construct the topological classification of NH phases with $C_n$ symmetries.

\subsection{Classification of $C_n$-symmetric NH Hamiltonians in class AI}\label{Sec:HOTK and rotation}

The presence of crystalline symmetries enriches the classification of topological phases~\cite{Robert-Jan1,Slager_Mesaros_Juricic_Zaanen_2012,Morimoto_2013,teo2013,benalcazar_classification_2014,Shiozaki_2014,Shiozaki_2016,Shiozaki_2017,benalcazar_quantization_2019,zak_berrys_1989,po_symmetry-based_2017,bradlyn_topological_2017,vaidya2023}. In 2D NH Hamiltonians, HOSE phases have been understood by drawing a correspondence between the NH Hamiltonian in question $h_\mathrm{NH}$ and an associated chiral-symmetric Hermitian Hamiltonian $h_\mathrm{H}=[0,h_\mathrm{NH};h^\dagger_\mathrm{NH},0]$~\cite{mudry1998,mudry1998_2}, both of which carry identical topological information~\cite{kawabata_higher-order_2020}. This correspondence then makes use of existing crystalline topological classifications of $h_\mathrm{H}$ to diagnose topological phenomena in $h_\mathrm{NH}$~\cite{Shiozaki_2021,denner2023,Vecsei_2021,Okugawa_2021,Congcong_2022,zhong2024,wang2024classifyingordertwospatialsymmetries}.

Here, we consider the crystalline classification of $h_\mathrm{NH}$ itself, i.e., without reference to its corresponding $h_H$. The utility of such a classification was recently presented in Ref.~\cite{zhong2024} for a $C_3$-symmetric NH Hamiltonian, and here we extend this classification to all $C_n$ symmetries. Specifically, consider NH Hamiltonians obeying TRS \eqref{eq:regular TRS} and $C_n$ symmetry,
\begin{align}
    \hat{r}_n h(\textbf{k})\hat{r}_n^{-1}=h( R_n\textbf{k}),
    \label{eq:CnSymmetry}
\end{align}
where $\hat{r}_n$ is the rotation operator acting on the unit cell degrees of freedom, obeying $[\hat{r}_n]^n=1$ or $[\hat{r}_n]^n=-1$ (the latter case due to, e.g., a magnetic flux threading the lattice), and $R_n$ is the $n$-fold rotation matrix acting on the crystal momentum $\textbf{k}$. The high symmetry points (HSPs) in the BZ zone, ${\bf \Pi}_m$, for $m\leq n$, are crystal momenta that remain invariant under the little-group $C_m$ rotation (modulo a reciprocal lattice vector ${\bf G}$), i.e., ${\bf \Pi}_m=R_m{\bf \Pi}_m$ (mod ${\bf G}$). For example, setting the length of the unit cell to unity, in $C_2$ symmetric lattices, ${\bf X}=(\pi,0)$, ${\bf Y}=(0,\pi)$, and ${\bf M}=(\pi,\pi)$ are all invariant under $C_2$ rotations, while in $C_4$ symmetric lattices, ${\bf X}=(\pi,0)$ and ${\bf X'}=(0,\pi)$ are $C_2$ invariant and ${\bf M}=(\pi,\pi)$ is $C_4$ invariant. Also, notice that trivially, ${\bf \Gamma}=(0,0)$ is invariant under the full group of a $C_n$ symmetric lattice.  Appendix~\ref{sec:ConstructionOfClassification} describes the HSPs for all $C_n$-symmetric lattices. 

At the HSPs, Eq.~\eqref{eq:CnSymmetry} implies that $[\hat{r}_m,h({\bf  \Pi}_m)]=0$ and thus $\hat{r}_m$ and $h({\bf  \Pi}_m)$ have simultaneous eigenstates; the Bloch eigenstates of energy band $l$, $\ket{u^l_{\bf \Pi_m}}$, which obey $h({\bf \Pi}_m)\ket{u^l_{{\bf \Pi}_m}}=\epsilon_l({\bf \Pi}_m)\ket{u^l_{{\bf \Pi}_m}}$, simultaneously obey
\begin{equation}
    \hat{r}_m\ket{u_{{\bf \Pi}_m}^l}=r_{{\bf \Pi}_m}^l\ket{u_{{\bf \Pi}_m}^l},
\end{equation}
where $r_{{\bf \Pi}_m}^l$ is the rotation eigenvalue associated with energy band $l$ at HSP ${\bf \Pi}_m$, which can take the values
\begin{equation}
    {\Pi}_p^{(m)}=
    \begin{cases}
        e^{2\pi \textrm{i}(p-1)/m}, & \text{for }[\hat{r}_n]^n=1\\
        e^{2\pi \textrm{i}(p-1/2)/m}, & \text{for }[\hat{r}_n]^n=-1
    \end{cases}
\end{equation}
for $p=1,2,\dots m$. Extending previous studies on the classification of crystalline topological phases~\cite{teo2013,benalcazar_classification_2014,benalcazar_quantization_2019,schindler2019,vaidya2023}, we define the symmetry indicator invariants for energy band $l$ as
\begin{equation}
    [{\Pi}_p^{(m)}]=\#_l{\Pi}_p^{(m)}-\#_l\Gamma_p^{(m)},
    \label{eq: crystalline symmetry invariant}
\end{equation}
where $\#_l{\Pi}_p^{(m)}$ is the number of eigenstates in the band $l$ with rotation eigenvalue ${\Pi}_p^{(m)}$ at HSP ${\bf \Pi}_m$. The set of symmetry indicator invariants \eqref{eq: crystalline symmetry invariant} across all HSPs of the BZ, along with the Chern number, provide a topological classification for the energy band $l$. However, some of these invariants are related to one another by symmetry, and thus there is redundancy in the topological information (Appendix~\ref{sec:ConstructionOfClassification}). We collect the set of non-redundant indicators in a vector index $\chi^{(n)}$ that uniquely identifies the topological class of complex energy bands in $C_n$-symmetric lattices (Appendix \ref{sec:ConstructionOfClassification} contains the complete derivation of the $\chi^{(n)}$ indices for all $C_n$-symmetric Bloch Hamiltonians). These $\chi^{(n)}$ indices are
\begin{align}
    \chi^{(2)} &=(C | [X_1^{(2)}],[Y_1^{(2)}],[M_1^{(2)}]),\nonumber\\
    \chi^{(4)} &=(C | [X_1^{(2)}],[M_1^{(4)}],[M_2^{(4)}],[M_3^{(4)}]),\nonumber\\
    \chi^{(3)} &=(C | [K_1^{(3)}],[K_2^{(3)}],[K_1^{'(3)}],[K_2^{'(3)}]),\nonumber\\
    \chi^{(6)} &=(C | [M_1^{(2)}],[K_1^{(3)}],[K_2^{(3)}]).
    \label{eq:IndicesComplexBands}
\end{align}

Non-Hermitian $C_n$-symmetric Hamiltonians in class AI with different $\chi^{(n)}$ indices belong to different topological phases, as they cannot be deformed into one another without closing the bulk energy gaps or breaking the symmetry.

The $\chi^{(n)}$ indices obey an algebraic structure. If two bands $a$ and $b$ in classes $\chi^{(n)}_{a}$ and $\chi^{(n)}_{b}$ are combined, the resulting Hamiltonian is in class $\chi^{(n)}_{a \cup b}=\chi^{(n)}_{a} + \chi^{(n)}_{b}$. As a consequence, if two bands have complementary topological indices, i.e., if $\chi^{(n)}_{a} = - \chi^{(n)}_{b}$, such that $\chi^{(n)}_{a \cup b}=0$, boundary states will exist in the gap between them under OBC. 

As an example, consider the Bloch Hamiltonian $h_\mathrm{HOSE}({\bf k})$ of Eq.~\eqref{eq:HOSE} for the minimal model of a HOSE phase. While $h_\mathrm{HOSE}({\bf k})$ does not obey TRS~\eqref{eq:regular TRS}, $\ii h_\mathrm{HOSE}({\bf k})$ does, with $\mathcal{T}=\mathcal{K}$. Additionally, $\ii h_\mathrm{HOSE}({\bf k})$ obeys $C_2$ symmetry, with rotation operator $\hat{r}_2=\sigma_y$.  The $\chi^{(2)}$ indices for its two bulk energy bands are shown in Appendix \ref{Appendix: chi2 index HOSE}. The two indices trivialize in pairs, leading to boundary states in the gap between these two bands under OBC. These are the skin effect modes of the HOSE phase. 

This pairwise trivialization of bands is the simplest case of a more general topological connectivity between bands. In particular, there are cases in which no pairwise trivialization occurs; instead, bands trivialize only in groups of three or four. As we will see, these cases can undergo unusual phase transitions in which the line gaps close, generating a point gap, from which edge-localized states with nontrivial windings emerge. These are the phases we refer to as ``higher-order topological knot" (HOTK) phases, for reasons to become apparent in the next section.

\subsection{Model Hamiltonians with HOTK phases}\label{Sec:HOTK}

Higher-order topological knot phases are NH topological phases with \emph{bulk line gaps} and nontrivial weak topology that generate edge states with \emph{edge point gaps} with nontrivial windings along all edges. Note that similar characteristics are exhibited by HOSE phases. However, in HOSE phases, nontrivial edge states exist only at certain edges. In Section~\ref{sec:DeformationComplexChern}, we showed that a two-band model with a line gap and nonzero Chern numbers can only transition into a HOSE phase because, while the chiral edge states deform into a pair of edge states with a nontrivial winding number in the complex energy plane, the other pair of chiral edges merge into the bulk, resulting in no topological states in the HOSE phase. This suggests that the smallest model that generates topological edge states along all edges requires more than two line-gap energy bands. 

\begin{figure}
    \centering
    \includegraphics[width=\columnwidth]{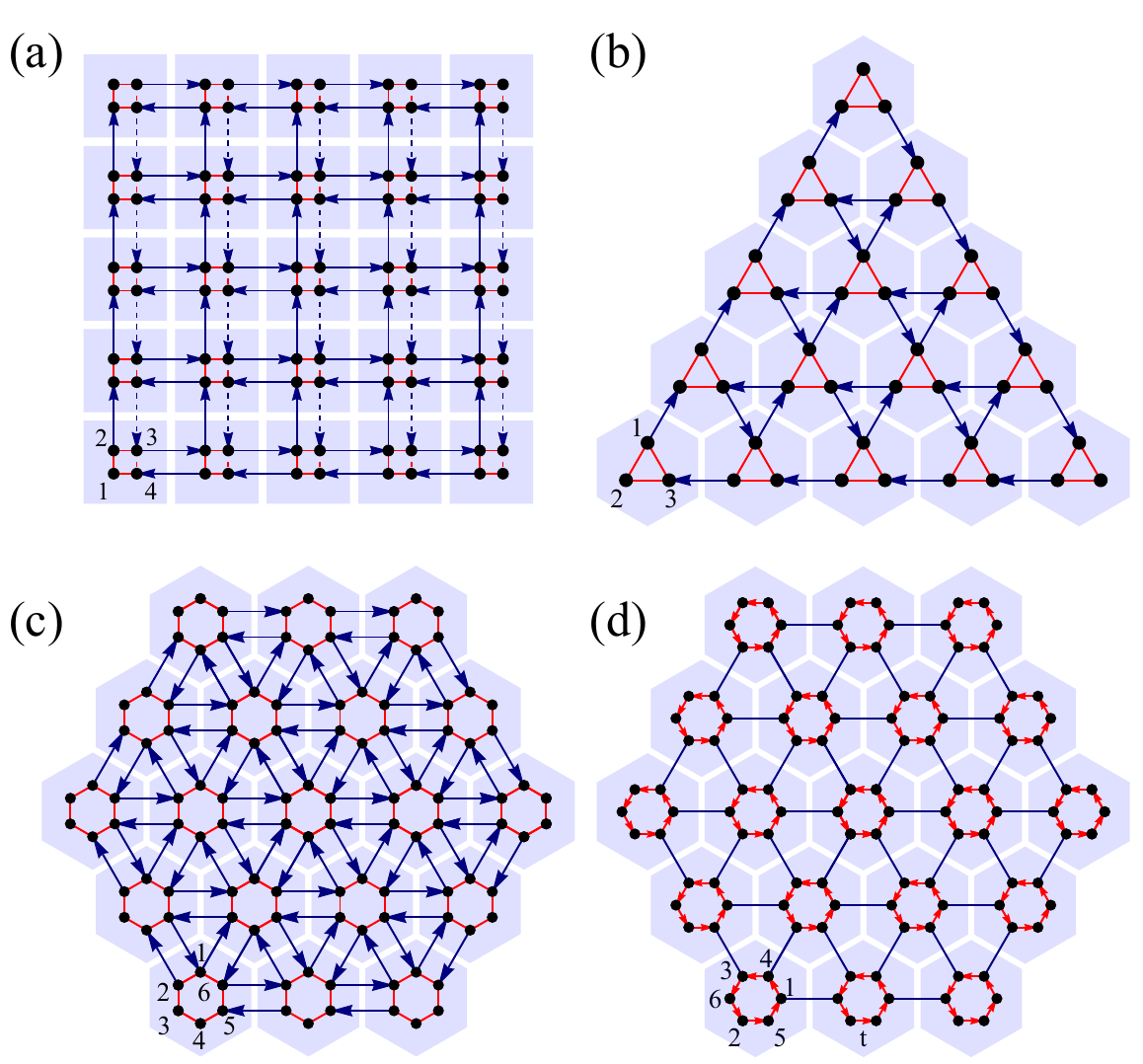}
    \caption{Tight-binding lattices that support various NH topological phases. Arrows indicate the direction of nonreciprocal hopping. (a) The $C_4$-symmetric lattice with the Bloch Hamiltonian given in Eq.~\eqref{eq:h4} hosts both a HOTK phase and an imaginary-line-gap complex Chern insulator phase. The hopping terms corresponding to dotted lines carry a $-1$ sign, a gauge choice to account for a $\pi$ flux per plaquette. (b,c) The $C_3$-symmetric ($C_6$-symmetric) lattice with the Bloch Hamiltonian in Eq.~\eqref{eq:h3} [Eq.~\eqref{eq:h6}], which hosts a HOTK phase. The $C_3$-symmetric model also supports an imaginary-line-gap complex Chern insulator phase. (d) The $C_6$-symmetric lattice with the Bloch Hamiltonian in Eq.~\eqref{eq:NH Breathing HoneyComb}, which supports a real-line-gap complex Chern insulator phase.}
    \label{fig:Lattices}
\end{figure}

In what follows, we will first present a four-band model for a $C_4$-symmetric lattice that exists in either a complex Chern insulator phase or in a HOTK phase. We will then describe HOTK models obeying $C_3$ and $C_6$ crystalline symmetries. Figure~\ref{fig:Lattices} illustrates the lattices we consider.
The Hamiltonians for these lattices are written in a ``maximally nonreciprocal limit'', where the arrows in Fig.~\ref{fig:Lattices} represent hoppings only along their directions. We do this to provide the simplest, basic realizations of these phases; however, more generally, the arrows can represent an imbalance in the amplitude of hoppings along their direction vs their opposite direction. The lines with no arrow represent reciprocal hoppings.

\subsubsection{Protected by $C_4$ symmetry}\label{sec:C4 symmetric model main text}
\begin{figure*}
    \centering
    \includegraphics[width=\textwidth]{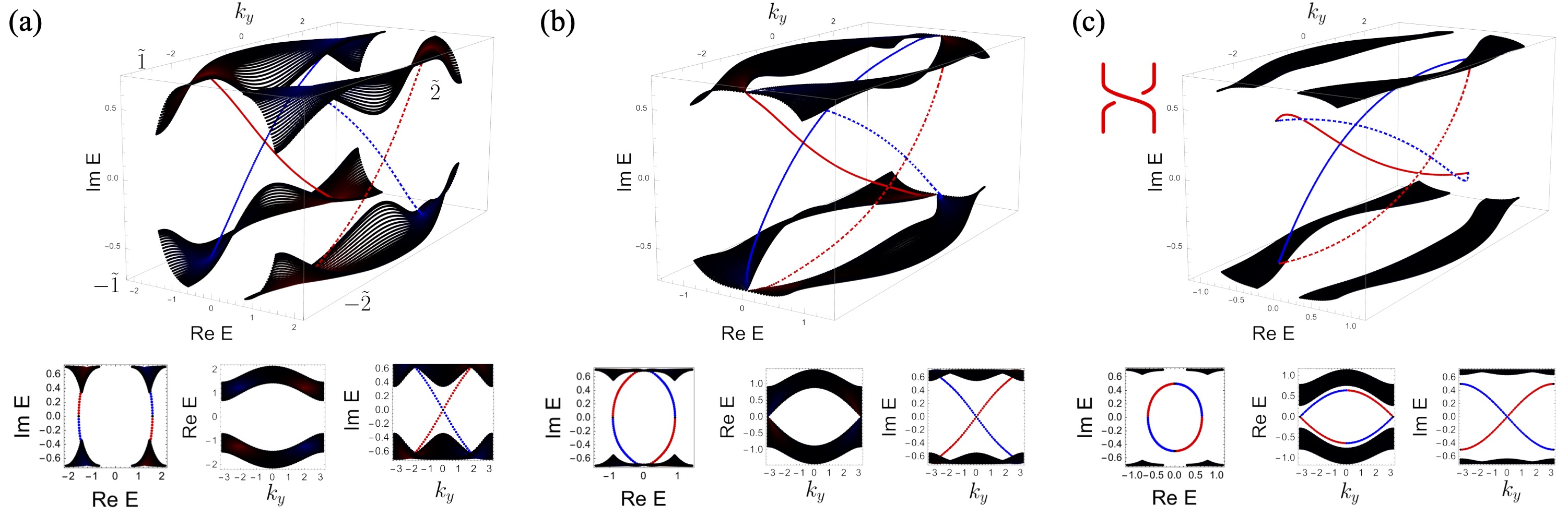}
    \caption{Energy spectra of Hamiltonian~\eqref{eq:h4} in the $(\text{Re} E, \text{Im} E, k)$ space under OBC along $x$ and PBC along $y$ with (a) $t=1$, (b) $t=0.5$, and (c) $t=0.3$. Because of $C_4$ symmetry, the energy spectra under OBC along $y$ and PBC along $x$ are identical to these plots. Black indicates bulk states; red and blue indicate states localized at opposite edges. The lower panels show projections of the 3D plots into 2D planes to aid visualization. The diagram at the top-left corner of (c) indicates the braid group of each of the edge states across the BZ.}
    \label{fig:h4}
\end{figure*}
Consider the following Bloch Hamiltonian, which hosts both a complex Chern insulator phase and a HOTK phase,
\begin{align}
h^{(4)}({\bf k}) = & t \sigma_x(\tau_x + \tau_z)  \nonumber\\
&+\frac{1}{2}(\sigma_x+\textrm{i}\sigma_y)[\tau_x\cos(gk_x)-\tau_y\sin(gk_x)]\nonumber\\
&+\frac{1}{2}(\sigma_x-\textrm{i}\sigma_y)[\tau_z\cos (gk_y)-\textrm{i}\tau_0\sin (gk_y)],
\label{eq:h4}
\end{align}
where both $\sigma_i$ and $\tau_i$, for $i=x,y,z$, are Pauli matrices and $\sigma_0$ and $\tau_0$ are the $2 \times 2$ identity matrices. 
This model is parametrized by $t\in\mathbb{R}^+$, the amplitude of reciprocal hoppings within the unit cell, and $g\in \mathbb{Z}^+$, the nonreciprocal hopping distance between unit cells. We first consider the case $g=1$, which corresponds to nearest-neighbor intercell hopping. The lattice of this model is shown in Fig.~\ref{fig:Lattices}(a) and carries a $\pi$ flux per plaquette, which is accounted for by a $-1$ sign in the hopping terms corresponding to dotted lines in Fig.~\ref{fig:Lattices}(a). This model resembles the quadrupole topological insulator model of Ref.~\cite{Benalcazar_Bernevig_Hughes_science_2017}, but with the crucial distinction that its intercell hopping terms are nonreciprocal. It obeys TRS~\eqref{eq:regular TRS} with $\mathcal{T}=\mathcal{K}$ and $C_4$ symmetry \eqref{eq:CnSymmetry} with rotation operator
\begin{align}
\hat{r}_4=\left(
\begin{array}{cccc}
 0 & 0 & 0 & 1 \\
 1 & 0 & 0 & 0 \\
 0 & 1 & 0 & 0 \\
 0 & 0 & -1 & 0 \\
\end{array}
\right), \nonumber
\end{align}
which obeys $\hat{r}^4_4=-1$. 

The Hamiltonian \eqref{eq:h4} has four bands, each occupying a quadrant on the complex plane. These four bands are labeled according to the rule introduced in Sec. \ref{Sec: complex Chern insulators} as shown in Fig.~\ref{fig:h4}(a).

When $t>0.5$, the Hamiltonian~\eqref{eq:h4} is in a complex Chern insulator phase. The  $\chi^{(4)}$ indices of each of the four bands in this phase are given in Table \ref{tab:chi4}. Note that bands $\Tilde{1}$ and $\Tilde{2}$ have complementary topologies to those of bands $-\Tilde{1}$ and -$\Tilde{2}$, respectively. Consequently, the topological edge states cross the imaginary line gap [Fig.~\ref{fig:h4}(a)].

\begin{table}
\begin{tabular}{lrrrrrrr}
  \toprule
  Phase &Band & $C$ & $[X_1^{(2)}]$ & $[M_1^{(4)}]$ & $[M_2^{(4)}]$ &  $[M_3^{(4)}]$& \\
  \midrule
\multirow{4}{*}{complex Chern} 
& $\Tilde{1}$ &-1  & 1     & 0         & -1       &      1&      \\
&$-\Tilde{1}$ &1   & -1    & 0         & 1        &     -1&      \\
&$\Tilde{2}$  &1   & -1    & -1        & 0        &      0&      \\
&$-\Tilde{2}$ &-1  & 1     & 1         & 0        &      0&      \\
\midrule
\multirow{4}{*}{HOTK}
&$\Tilde{1}$  &0 & 1      & 0         & -1       &      0&       \\
&$-\Tilde{1}$ &0 & -1     & 1         & 0        &     -1&       \\
&$\Tilde{2}$  &0 & -1     & -1        & 0        &      1&       \\
&$-\Tilde{2}$ &0 & 1      & 0         & 1        &      0&       \\
\bottomrule
\end{tabular}
\caption{$\chi^{(4)}$ indices for the energy bands of Hamiltonian~\eqref{eq:h4}. Bands are labeled as indicated in Fig.~\ref{fig:h4}(a). The complex Chern insulator and HOTK phases correspond to $t>0.5$ and $t<0.5$, respectively.}
\label{tab:chi4}

\begin{tabular}{lrrrrrr}
\\
    \toprule
    Phase &Band &  $C$ & $[X_1^{(2)}]$ & $[Y_1^{(2)}]$ & $[M_1^{(2)}]$ & \\ \hline
    \multirow{4}{*}{complex Chern}
    &$\Tilde{1}$  &-1 & 1  & 1  & 1  &   \\
    &$-\Tilde{1}$ &1  & -1 & -1 & -1 &   \\
    &$\Tilde{2}$  &1  & -1 & -1 & -1 &   \\
    &$-\Tilde{2}$ &-1 & 1  & 1  & 1  &   \\
    \midrule
    \multirow{4}{*}{HOTK}
    &$\Tilde{1}$  &0 & 1  & 1  & 0 &    \\
    &$-\Tilde{1}$ &0 & -1 & -1 & 0 &    \\
    &$\Tilde{2}$  &0 & -1 & -1 & 0 &    \\
    &$-\Tilde{2}$ &0 & 1  & 1  & 0 &    \\
    \bottomrule
\end{tabular}
\caption{$\chi^{(2)}$ indices for the energy bands of Hamiltonian~\eqref{eq:h4}.}
\label{tab:chi2}
\begin{tabular}{lrrrrrrr}
\\
\toprule
    Phase& Band& $C$ & $[K_1^{(3)}]$  & $[K_2^{(3)}]$  &  $[K_1^{'(3)}]$ & $[K_2^{'(3)}]$ & \\
    \midrule
    \multirow{3}{*}{HOTK}
    &$\Tilde{1}$  &0  & 0   & -1  &1   &-1& \\
    &$-\Tilde{1}$ &0  & 1   & 0   &0   & 1& \\
    &$\Tilde{2}$  &0  & -1  & 1   &-1  &0 & \\
    \midrule
    \multirow{4}{*}{complex Chern}
    &$\Tilde{1}$  &2  & 0  & -1  &0  &-1 & \\
    &$-\Tilde{1}$ &-2 & 0  & 1   &0  &1  &  \\
    &$\Tilde{2}$  &0  & 0  & 0   &0  &0  & \\
    \bottomrule
\end{tabular}
\caption{$\chi^{(3)}$ indices for Hamiltonian \eqref{eq:h3}. The HOTK and complex Chern insulator phases correspond to $0<t<0.57$ and $0.57<t<1$, respectively.}
\label{tab:chi3}
\begin{tabular}{lrrrrrr}
\\
\toprule
    Phase& Band& $C$ & $[M_1^{(2)}]$  & $[K_1^{(3)}]$  &  $[K_2^{(3)}]$&  \\
    \midrule
    \multirow{5}{*}{HOTK}
    &$\Tilde{1}$  &0  &-1  &0  &-1 &\\
    &$\Tilde{2}$  &0  &1   &1  &-1 &\\
    &$\Tilde{3}$  &0  &0   &-2 &1  & \\
    &$-\Tilde{1}$ &0  &1   &1  &1  & \\
    &$-\Tilde{2}$ &0  &-1  &0  &0  & \\
    \bottomrule
\end{tabular}
\caption{$\chi^{(6)}$ indices for Hamiltonian~\eqref{eq:h6} in the HOTK phase, which corresponds to $0<t<0.5$.}
\label{tab:chi6}
\end{table}

A transition to a trivial phase would close the imaginary line gap to connect band $\Tilde{1}$ with band -$\Tilde{1}$, and band $\Tilde{2}$ with band -$\Tilde{2}$, so that on the other side of the transition, $\chi^{(4)}={\bf 0}$ for all four bands, causing the edge states to disappear. However, a different phase transition occurs in this model when $t=0.5$. This transition closes the real line gap instead, and connects band $\Tilde{1}$ with band $\Tilde{2}$ and band $-\Tilde{1}$ with band $-\Tilde{2}$. Along with these bulk gap closings, the edge states also close the real-line gap [Fig.~\ref{fig:h4}(b)].

When $t<0.5$, that is, on the other side of the nontrivial transition, the Hamiltonian~\eqref{eq:h4} is in a HOTK phase [Fig.~\ref{fig:h4}(c)]. The four bulk energy bands present real and imaginary line gaps, as before the transition. However, edge states, now separated from the bulk bands, braid across the BZ in the complex energy plane around $E_p=0$. Since there are two edge bands for each edge, the corresponding braid group is $B_2$. The braid structure for one of the edge states is schematically illustrated at the top left corner of Fig.~\ref{fig:h4}(c). Because of the periodicity of the BZ, the braid is also a knot, and for this phase, the knot has periodicity $4\pi/a$, where $a$ is the unit-cell length, i.e., it takes going around the BZ twice to get back to the original point in the edge spectrum. The plots of energy bands for OBC along $y$ and PBC along $x$ (not shown) are identical to those in Fig.~\ref{fig:h4}, by $C_4$ symmetry. Thus, topological edges exist along the entire 1D boundary of the HOTK phase in Hamiltonian \eqref{eq:h4}. 

\begin{figure}
    \centering
    \includegraphics[width=\columnwidth]{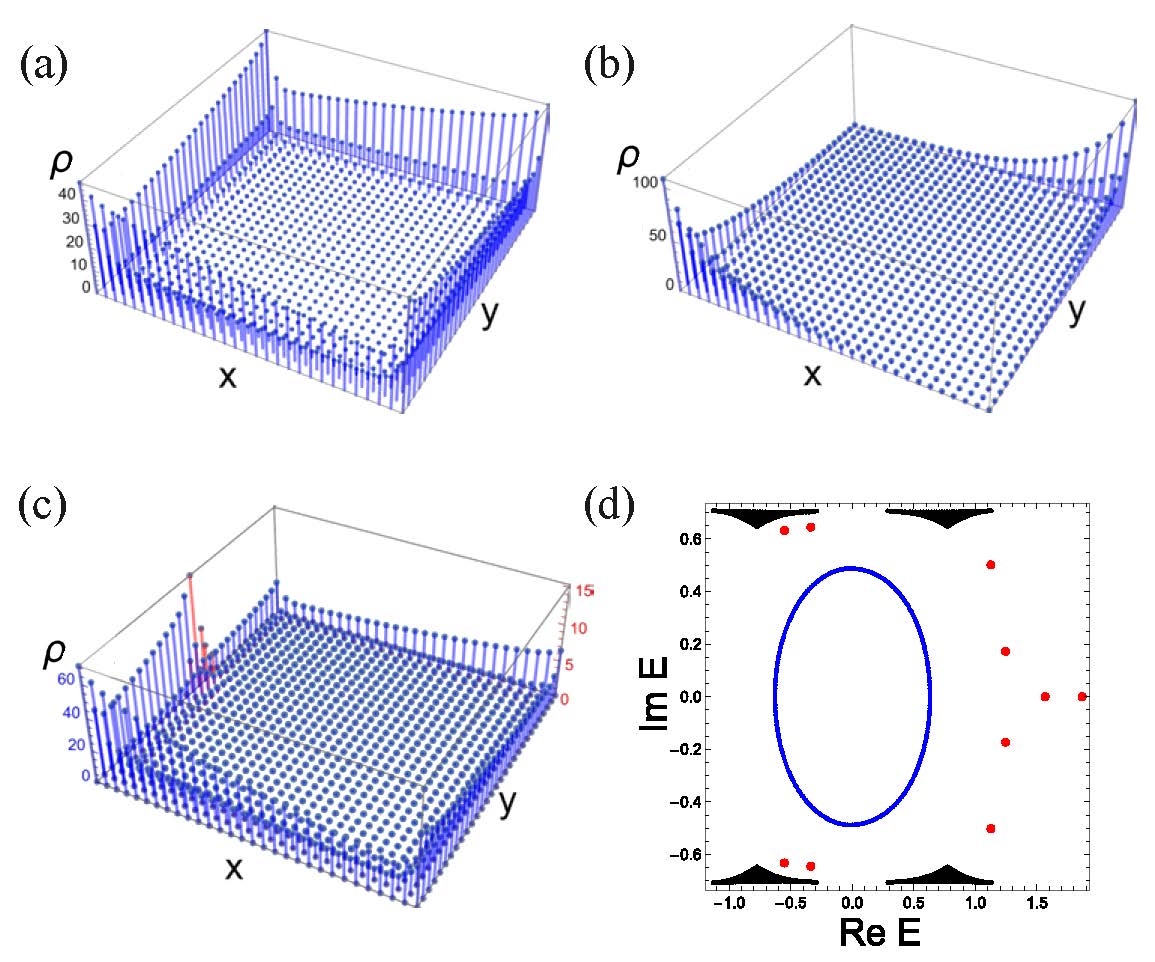}
    \caption{Local density of edge states $\rho$ in the HOTK phase of Hamiltonian \eqref{eq:h4} under full OBC ($30 \times 30$ unit cells) for three scenarios: (a) when $C_4$ symmetry is preserved, (b) when only $C_2$ symmetry is preserved by reducing hopping amplitudes along the $y$ direction, and (c) when a defect is added to one edge. In (b), the $C_4$ symmetry of Hamiltonian \eqref{eq:h4} has been reduced to only $C_2$ symmetry by setting the intercell hopping along $x$ to be $1$ and the one along $y$ to be 0.7, while setting the intracell hopping to $t=0.3$. Panel (d) shows the spectrum corresponding to case (c) where the defect is present. In both the plots of the energy spectrum and local density of states, edge states are indicated in blue, while defect states are highlighted in red.}
    \label{fig:Edge state localization}
\end{figure}

A plot of the density of edge states under full OBC for the HOTK phase is shown in Fig~\ref{fig:Edge state localization}(a). Notably, the states have support along all edges. This is significantly different from the phenomenon of HOSE, in which eigenstates decay exponentially from corners~\cite{kawabata_higher-order_2020}. In the bulk bands, the nontrivial topology manifests in the nonzero $\chi^{(4)}$ indices shown in Table~\ref{tab:chi4}. Note that none of these bands can be trivialized in pairs; only the $\chi^{(4)}$ index for the ensemble of all four bands is trivial. This indicates that all four bands participate in the nontrivial topology of the HOTK phase.

The boundary states essentially constitute a boundary-localized 1D chain with PBC and an associated non-zero winding number protected by a point gap at $E_p=0$, and thus, it lies in the same topological class as the 1D Hatano-Nelson model~\cite{Hatano1,Hatano2,HATANO3}. As such, the HOTK phase protects the nonreciprocal transport of wave packets along its 1D boundary, with unidirectional acceleration and amplification~\cite{PhysRevB.105.245143,Song_2019,PhysRevB.108.125402,PhysRevB.105.165137}.
\\~\\
\emph{Breaking $C_4$ symmetry and robustness of edge states.} The complex Chern insulator phase of Hamiltonian \eqref{eq:h4} does not require crystalline symmetries; even in the absence of $\chi^{(4)}$ indices, the nonzero Chern numbers $C=\pm1$ persist as long as the imaginary line gap is maintained. However, that is not the case for the HOTK phase, where all the bands have $C=0$. It may be tempting to think that $C_4$ symmetry is crucial for protecting the topological phase. However, the nontrivial $\chi^{(4)}$ indices of the HOTK phase are sufficient, but not necessary, to protect the phase. Boundary states will persist as long as the bulk line gaps and the gap between bulk and boundary states remain open. In the absence of $C_4$ symmetry, however, the localization of boundary states may significantly change. 

To examine the fate of the edge states in the HOTK phase, we break $C_4$ symmetry in three ways: (i) by making the hopping amplitude along $y$ different from that along $x$, (ii) by adding onsite energy perturbations $\epsilon \text{diag}(1,-1,1,-1)$ at each unit cell, and (iii) by adding a local onsite energy defect at an arbitrary point along the edge. While (i) and (ii) reduce $C_4$ symmetry to $C_2$ symmetry, (iii) breaks all global crystalline symmetries. The edge states for some of these cases are shown in Fig.~\ref{fig:Edge state localization}. In case (i), the edge states do not extend along the edges anymore and instead exponentially localize at corners, as shown in Fig.~\ref{fig:Edge state localization}(b). In case (ii), although $C_4$ symmetry is broken, the resulting distribution of edge states is similar to that of Fig.~\ref{fig:Edge state localization}(a). In case (iii), as shown in Fig.~\ref{fig:Edge state localization}(c), the defect does affect the distribution of edge states, localizing some of them, but leaving most of them distributed along the entire boundary. This shows that HOTK phases are robust against disorder: while the distribution of the boundary modes may change, they generally exist across all boundaries. Under only $C_2$ symmetry, the $\chi^{(2)}$ indices of the bulk bands are shown in Table~\ref{tab:chi2}. Note that the bands can now be trivialized in pairs across either the real line gap or the imaginary line gap; thus, transitions that close either of the two line gaps could unwind the edge states or make them disappear. \\~\\

\begin{figure}
    \centering
    \includegraphics[width=\columnwidth]{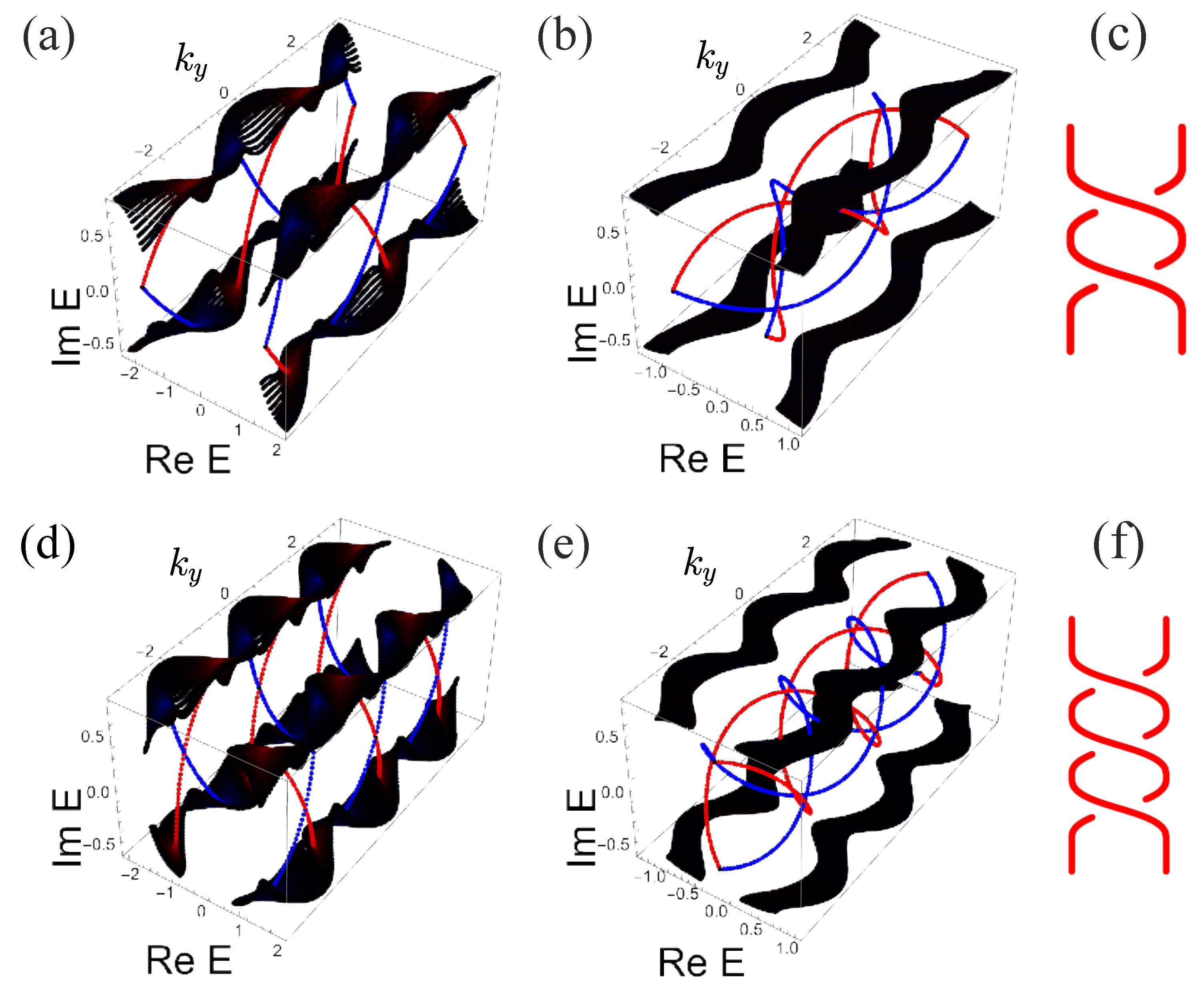}
    \caption{Energy spectra of Hamiltonian~\eqref{eq:h4} under OBC along $x$ and PBC along $y$ for $g=2$ (first row) and $g=3$ (second row). (First column) Complex Chern insulator phases with $t=1$. Each bulk band has a Chern number of magnitude $g^2$, specifically, $C=\pm4$ in (a) and $C=\pm9$ in (d). The edge states in (a) and (d) are twofold and threefold degenerate, respectively. (Second column) HOTK phases, with $t=0.3$. (Third column) Braid structure of the edge states at each edge of the HOTK phases.}
    \label{fig:Braid Generalization}
\end{figure}

\emph{Increasing the braiding of edge states.}
The braiding structure of the edge states across the BZ in the HOTK phase of Hamiltonian \eqref{eq:h4} is modified with increasing hopping distance $g$. For $g=2$, the lattice has only next-nearest neighbor hoppings between unit cells. The spectra for the complex Chern insulator phase and the HOTK phase are shown in Fig.~\ref{fig:Braid Generalization}(a) and \ref{fig:Braid Generalization}(b), respectively. Each band in the complex Chern insulator phase has $C=\pm4$, and each edge state in Fig.~\ref{fig:Braid Generalization}(a) is twofold degenerate. At each edge in the HOTK phase, the edge states braid forming a Hopf link, as shown in Fig.~\ref{fig:Braid Generalization}(c) for only one edge.
Similarly, setting $g=3$ fixes the Chern number of each band in the complex Chern insulator phase to $C=\pm9$, with each edge state in Fig.~\ref{fig:Braid Generalization}(d) being threefold degenerate. At each edge of the HOTK phase, the edge states braid forming a Trefoil knot, as shown in Fig.~\ref{fig:Braid Generalization}(f) for one of the edges.

\subsubsection{Protected by $C_3$ symmetry}
\begin{figure}
    \centering
    \includegraphics[width=\columnwidth]{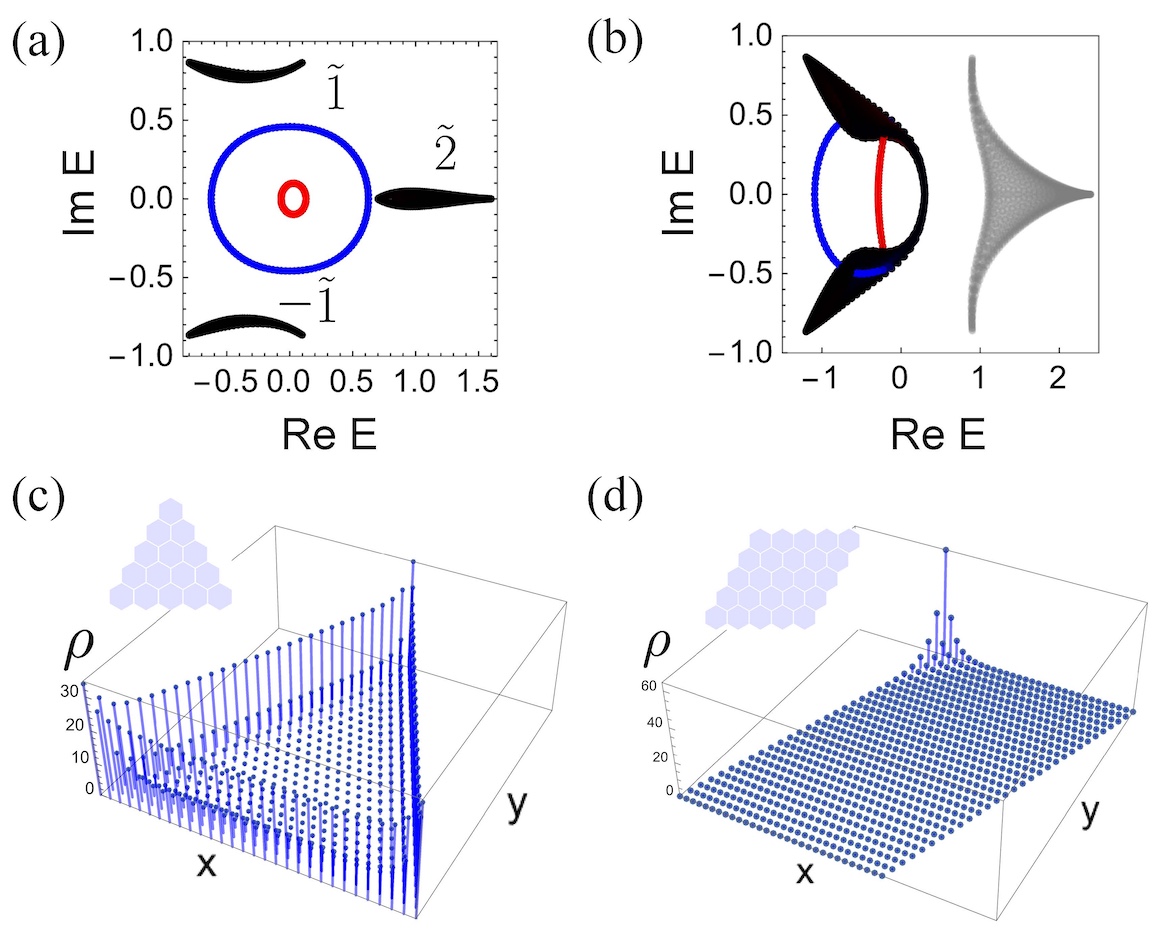}
    \caption{Energy spectra under OBC along $y$ and PBC along $x$ (a,b) and edge density of states under full OBC (c,d) of Hamiltonian~\eqref{eq:h3}. (a,c,d) are in the HOTK phase, with $t=0.3$. (b) is in the complex Chern insulator phase, with $t=0.7$. The insets in (c) and (d) schematically indicate the shape of the lattice.}
    \label{fig:h3}
\end{figure}

We now consider the NH Kagome lattice shown in Fig.~\ref{fig:Lattices}(b). It has Bloch Hamiltonian
\begin{equation}
    h^{(3)}({\bf k})=\left(
\begin{array}{ccc}
 0 & t & t+e^{-i \textbf{k}\cdot \textbf{a}_3} \\
 t+e^{-i \textbf{k}\cdot \textbf{a}_2} & 0 & t \\
 t & t+e^{i\textbf{k}\cdot \textbf{a}_1} & 0 \\
\end{array}
\right),
\label{eq:h3}
\end{equation}
 where $\textbf{a}_1=(1,0),\textbf{a}_2=(1/2,\sqrt{3}/2),\textbf{a}_3=(1/2,-\sqrt{3}/2)$. This model obeys TRS~\eqref{eq:regular TRS} with $\mathcal{T}=\mathcal{K}$ and $C_3$ symmetry with rotation operator
 \begin{align}
    \hat{r}_3=\left(
\begin{array}{ccc}
 0 & 0 & 1 \\
 1 & 0 & 0 \\
 0 & 1 & 0  \\
\end{array}
\right). \nonumber
\end{align}
For $0<t<0.57$, this model is in a HOTK phase. Its spectrum at $t=0.3$ is shown in Fig.~\ref{fig:h3}(a). The $\chi^{(3)}$ index in this phase is shown in Table \ref{tab:chi3}, where we have labeled the bands as indicated in Fig.~\ref{fig:h3}(a). Note that no pairs of bands lead to a trivial index; a trivial $\chi^{(3)}$ index is obtained only for the set of all three bands, as expected for a HOTK phase. Under full OBC, a crystal that preserves $C_3$ symmetry, such as the triangular crystal shown in Fig.~\ref{fig:Lattices}(b), hosts point-gapped edge states along all edges, as shown in Fig.~\ref{fig:h3}(c). These edges are then associated with nonreciprocal propagation.

When a crystal under full OBC breaks $C_3$ symmetry, the edge states can collapse into corners. This phenomenon was recently presented in Ref.~\cite{zhong2024} as realizing a HOSE phase in the Kagome lattice. The exponential localization of states to a corner is similar to what we found in the HOTK phase of Hamiltonian~\eqref{eq:h4} when $C_4$ symmetry is reduced to $C_2$ symmetry [Fig.~\ref{fig:Edge state localization}(b)]. However, the corner states in the configuration of Fig.~\ref{fig:h3}(d) were demonstrated to have a correspondence with the zero-energy corner states of a chiral-symmetric Hermitian higher-order topological insulator, predicted and demonstrated experimentally in Ref.~\cite{HOTI_waveguides}.

For $0.57<t<1$, the Hamiltonian~\eqref{eq:h3} is in a complex Chern insulator phase with the $\chi^{(3)}$ indices shown in Table~\ref{tab:chi3}. As expected, the $\chi^{(3)}$ indices of bands $-\Tilde{1}$ and $\Tilde{1}$ are complementary in this phase, which necessarily leads to a trivial $\chi^{(3)}$ index for band $\Tilde{2}$. The energy spectrum for $t=0.7$ is shown in Fig.~\ref{fig:h3}(b). The chiral edge states cross the imaginary line gap between bands $-\Tilde{1}$ and $\Tilde{1}$. Although no direct line gap exists between $-\Tilde{1}$ and $\Tilde{1}$, these two bands are indeed separated everywhere in the BZ (Appendix~\ref{appendix:C3 and C6 spectrum}). 

\subsubsection{Protected by $C_6$ symmetry}\label{Sec:NH Hexagon Model}
Finally, we consider the NH Hamiltonian for the hexagonal lattice in Fig.~\ref{fig:Lattices}(c), described by the Hamiltonian
\begin{equation}
    h^{(6)}({\bf k})=\left(
\begin{array}{cccccc}
 0 & t & 0 & 0 & e^{-i \textbf{k}\cdot\textbf{a}_3} & t
   \\
 t & 0 & t & 0 & 0 & e^{-i \textbf{k}\cdot\textbf{a}_1} \\
 e^{-i \textbf{k}\cdot \textbf{a}_2} & t & 0 & t & 0 & 0
   \\
 0 & e^{i \textbf{k}\cdot\textbf{a}_3} & t & 0 & t & 0
   \\
 0 & 0 & e^{i \textbf{k}\cdot\textbf{a}_1} & t & 0 & t \\
 t & 0 & 0 & e^{i \textbf{k}\cdot\textbf{a}_2} & t & 0
   \\
\end{array}
\right).
\label{eq:h6}
\end{equation}
This model obeys TRS \eqref{eq:regular TRS} and $C_6$ symmetry with a rotation operator $\hat{r}_{6}$ that permutes the sites within the unit cells of the lattice in Fig.~\ref{fig:Lattices}(c) upon rotation by $2\pi/6$ about the center of the unit cell. 

\begin{figure}
    \centering
    \includegraphics[width=\columnwidth]{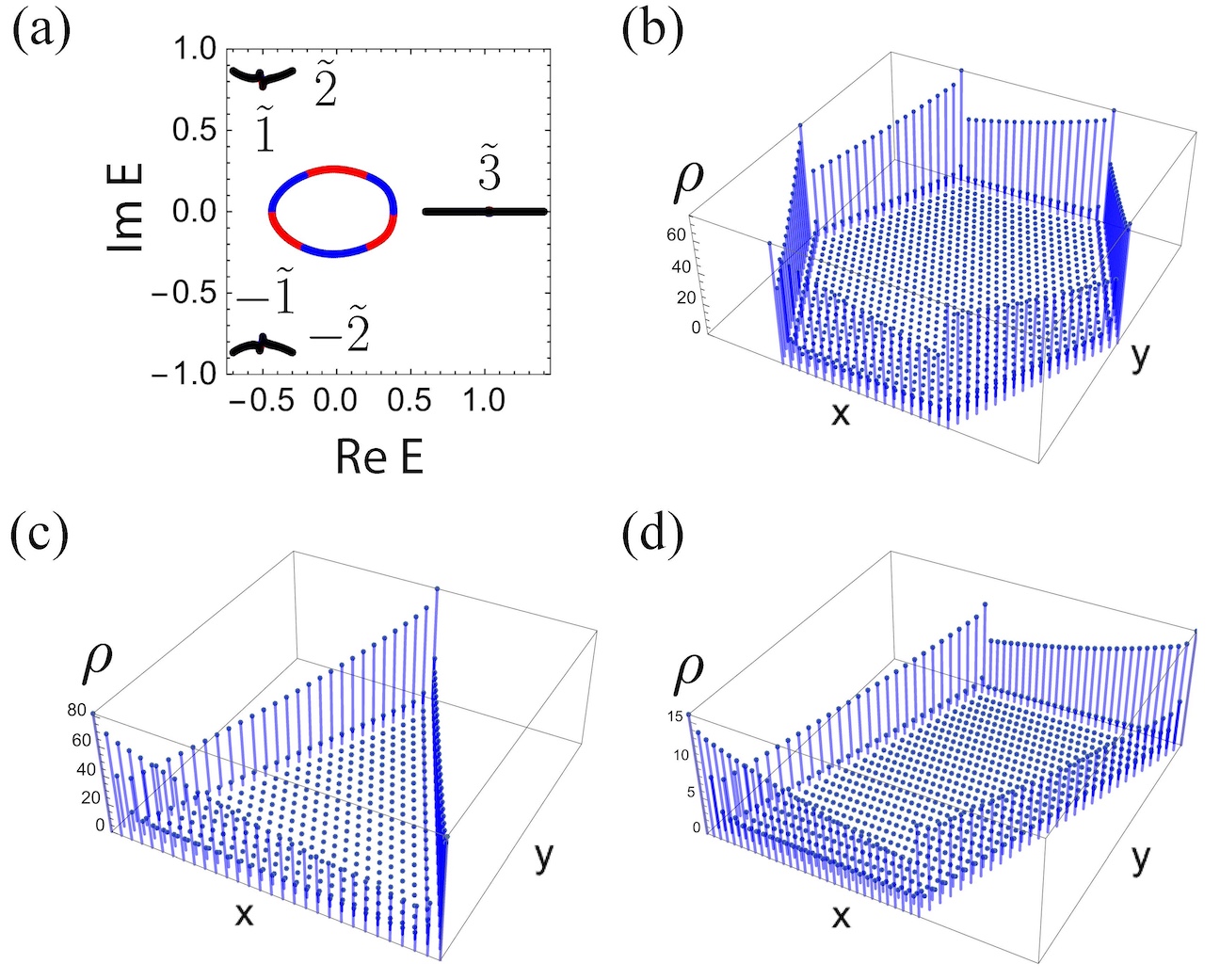}
    \caption{Energy spectra under PBC along $x$ and OBC along $y$ with a zigzag edge (a) and edge density of states under full OBC (b)-(d) of Hamiltonian~\eqref{eq:h6} in the HOTK phase with $t=0.2$.}
    \label{fig:h6}
\end{figure}

In the range $0<t<0.5$, the Hamiltonian \eqref{eq:h6} is in a HOTK phase. A phase transition happens at $t=0.5$. At $0.5<t<1$, Eq. \eqref{eq:h6} enters a gapless phase in which edge states merge into the bulk. A real line gap reopens at $t>1$. However, all $t>1$ phases are trivial. The $\chi^{(6)}$ indices for the HOTK phase are shown in Table~\ref{tab:chi6} [we labeled each band according to Fig.~\ref{fig:h6}(a)].

Figure~\ref{fig:h6}(a) shows the energy spectrum of Hamiltonian~\eqref{eq:h6} in the HOTK phase, at a value of $t=0.2$. Once again, none of the $\chi^{(6)}$ indices for each band can be trivialized in pairs or in any combination other than the one involving all bands. Under full OBC that preserves $C_6$ symmetry, such as the hexagon shown in Fig.~\ref{fig:Lattices}(c), edge states with point gaps and nontrivial winding exist along all edges [Fig.~\ref{fig:h6}(b)]. In fact, these states persist even in lattices that preserve only $C_3$ symmetry [Fig.~\ref{fig:h6}(c)] or $C_2$ symmetry [Fig.~\ref{fig:h6}(d)]. We show a more detailed spectrum of Hamiltonian~\eqref{eq:h6} in Appendix~\ref{appendix:C3 and C6 spectrum}. 

\section{Discussion and conclusions}\label{sec:discussion and conclusion}
We have introduced and characterized ``higher-order topological knot'' (HOTK) phases. These are NH phases defined by bulk bands with line gaps that support edge states with point gaps and nontrivial winding numbers. While these spectral characteristics are present in HOSE phases, there are two crucial distinctions: (i) in HOSE phases, the point-gap boundary states are always localized at corners; and (ii) these corner-localized states are skin states arising from nontrivial point-gap topology present only at certain edges of the lattice. In contrast, HOTK phases present boundary point-gap states extended across all edges, i.e., with no domain walls separating trivial and nontrivial edges, and thus exhibiting no skin effect.

In this paper, we focused on HOTK phases protected by $C_n$ symmetries. As markers of their topology, we identified symmetry indicator invariants for class AI under $C_n$ symmetry. Specifically, HOTK phases are characterized by bands with nonzero $\chi^{(n)}$ indices that cannot be trivialized in pairs, indicating that these phases arise from multi-band topology involving three or more bands, all separated by bulk line gaps. This topological connectivity is unique to NH systems, as multiple line band gaps are required to support it. 

In a specific example, we have shown how, once the phase has been established under a pristine $C_n$ symmetry, disorder that breaks the symmetry will not disrupt the existence of the topological states (unless disorder is strong enough to induce a bulk phase transition). We have also shown how breaking the protecting symmetries of a HOTK phase can result in $O(L)$ corner states, similar to those in HOSE phases. In this regard, it is worth distinguishing two cases: First, in Hamiltonian~\eqref{eq:h4}, reducing $C_4$ symmetry to $C_2$ symmetry results in $O(L)$ corner states that do not have corresponding zero-energy corner states in its counterpart chiral-symmetric Hermitian Hamiltonian, and the mechanism behind their localization remains an open question for future work. Second, in Hamiltonian~\eqref{eq:h3}, modifying the lattice geometry to break $C_3$ symmetry --- by shifting from a triangular to a parallelogram shape --- transforms the extended edge states into $O(L)$ corner states, localized exclusively at one of the two $2\pi/3$ rad corners. In this second case, the $O(L)$ states do correspond to zero-energy corner states in its counterpart chiral-symmetric Hermitian Hamiltonian, as recently identified in Ref.~\cite{zhong2024}. We point out, however, that Hamiltonian~\eqref{eq:h3} possesses edge point gaps along all edges under partial PBC, which contrasts with the paradigmatic HOSE phase of Ref.~\cite{kawabata_higher-order_2020}, where the corner states owe their localization to the absence of point-gap topology at one of the two pairs of opposite edges. 

In 1D NH Hamiltonians, point-gap topology can be characterized by the braid group $B_N$, which captures the winding of energy bands around each other in the complex energy plane as the BZ is traversed~\cite{hu_knots_2021}. The periodicity of the BZ allows these braids to be identified with knots. By analogy with higher-order topological insulators in Hermitian systems, where edge-state topology is determined by 2D bulk bands~\cite{Benalcazar_Bernevig_Hughes_science_2017}, we refer to our phases as ``higher-order topological knot'' phases, reflecting the fact that the point-gap knot topology of the 1D edge states is governed by the 2D bulk bands of the Hamiltonian. In this paper, the HOTK phases we have presented produced the edge braids and knots listed in Table~\ref{Table:knots and braids}.
A key task for future research is to understand how different classes of edge knots influence the nonreciprocal dynamics in these systems. 

\begin{table}[]
\begin{tabular}{ccccc}
\toprule
Name & Unlink & Hopf link & Trefoil knot & (3,2)Torus knot \\
\midrule
Braids & \includegraphics[scale=0.4]{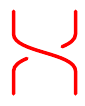}& \includegraphics[scale=0.3]{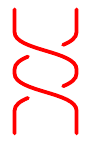}&  \includegraphics[scale=0.23]{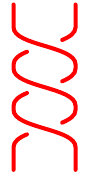}&   \includegraphics[scale=0.13]{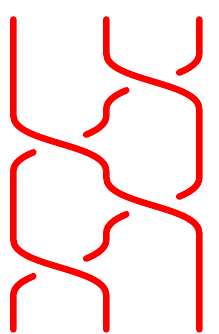}\\
BWs & $\sigma_1$ & $\sigma_1^2$ & $\sigma_1^3$ & $\sigma_1\sigma_2\sigma_1\sigma_2$\\
Knots & \includegraphics[scale=0.05]{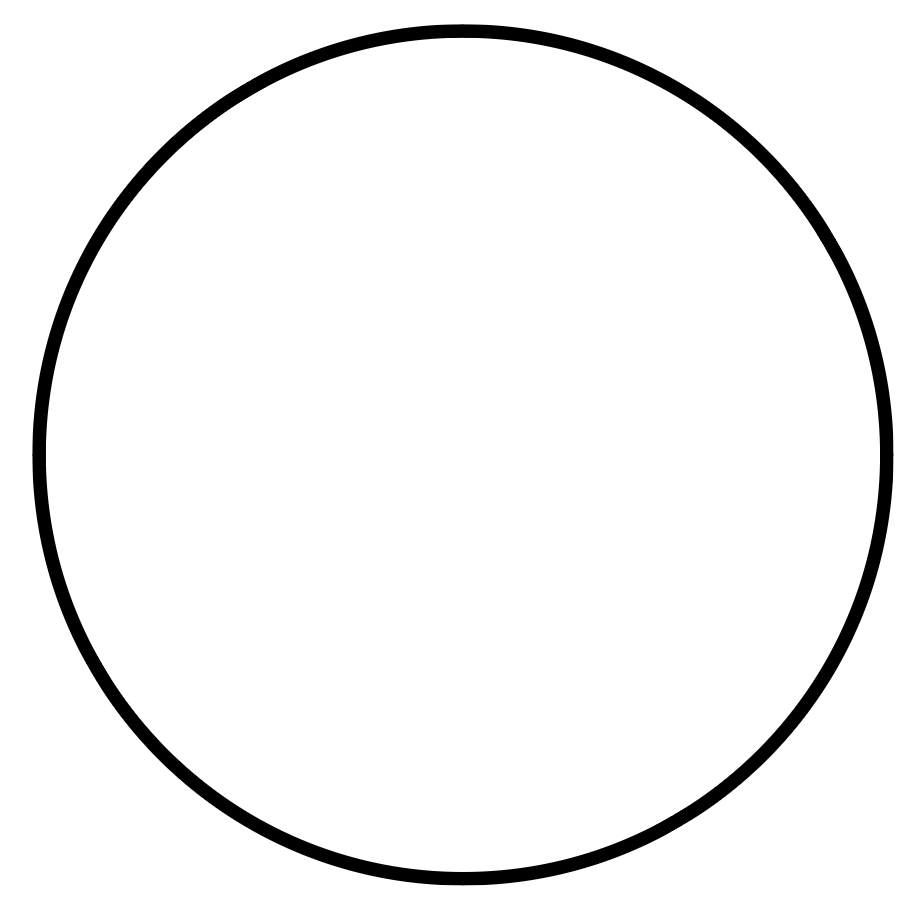}&  \includegraphics[scale=0.055]{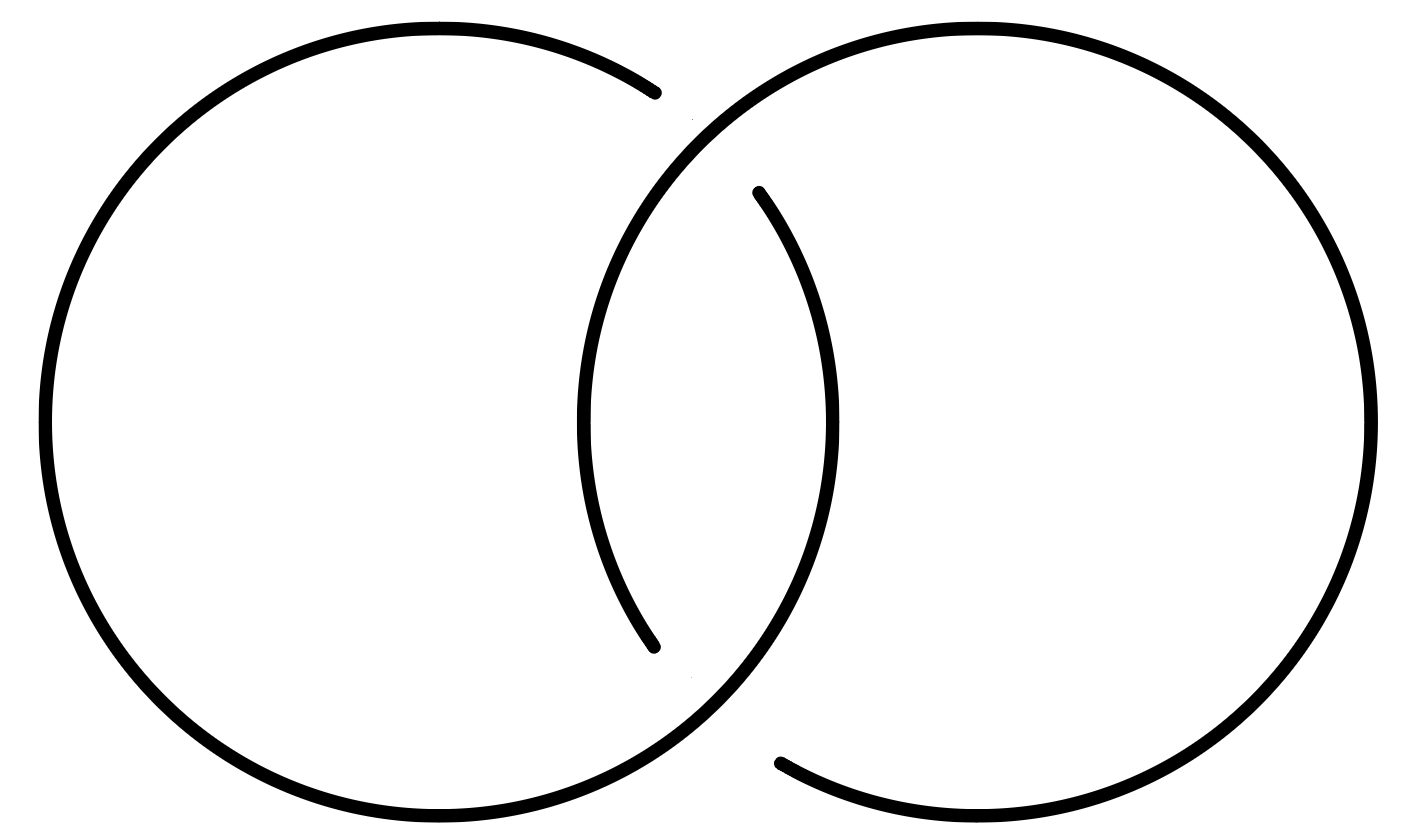}&  \includegraphics[scale=0.08]{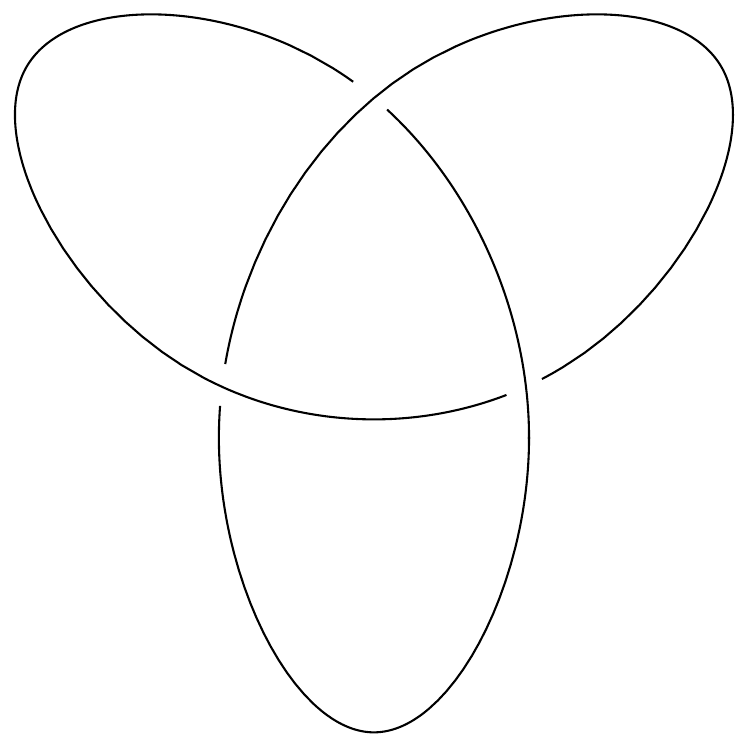}&   \includegraphics[scale=0.03]{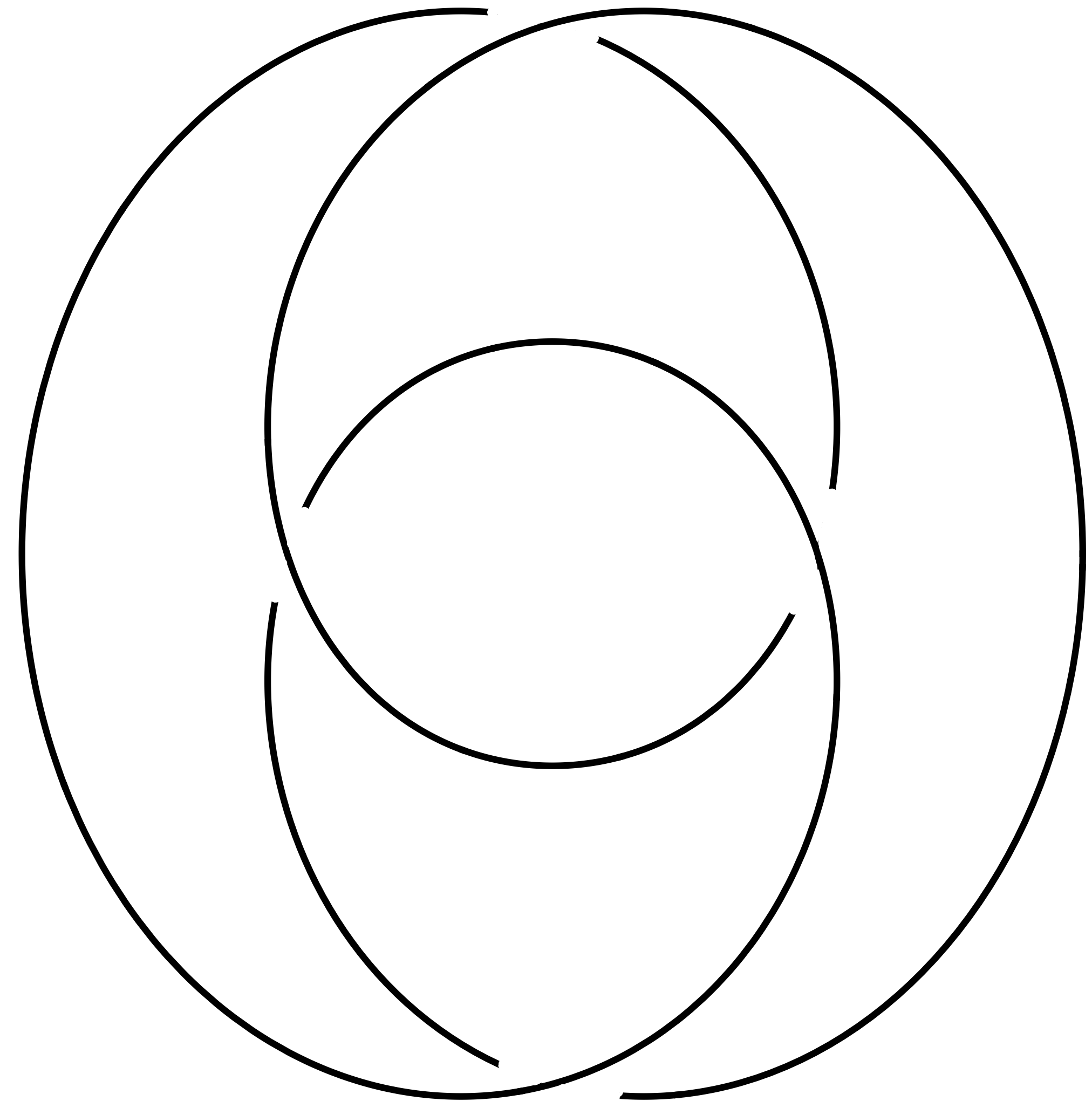}\\
Models & \eqref{eq:h4} [$g=1$] & \eqref{eq:h4} [$g=2$], \eqref{eq:h3} & \eqref{eq:h4} [$g=3$] & \eqref{eq:h6}     
\\
\bottomrule
\end{tabular}
\caption{Braids, braid words (BWs), and corresponding knots of the edge states in the HOTK phases described in Sec.~\ref{Sec: HOTK edge current}.}
\label{Table:knots and braids}
\end{table}

\section{Acknowledgments}
We thank Pedro Fittipaldi de Castro and Tianhong Lu for fruitful discussions on related topics. W.A.B. thanks the support of the startup funds from Emory University.


\appendix

\section{Winding number under TRS and TRS$^\dag$}\label{Appendix:winding number and TRS}
In this appendix, we demonstrate that in 1D, the winding number vanishes under TRS$^\dag$ \eqref{eq:pseudo TRS}, while it does not under TRS \eqref{eq:regular TRS}. To simplify our notation, we use $\mathcal{T}_a$ ($\mathcal{T}_b$) for the TRS$^\dag$ (TRS) operator in this appendix. 

First consider an 1D Hamiltonian that obeys TRS$^\dag$,
\begin{align}
    h(-k)^\dag&=\mathcal{T}_a h(k) \mathcal{T}_a^{-1} \nonumber\\
    &=U_a h(k)^* U_a^\dag,
\end{align}
where we used $\mathcal{T}_a=U_a \mathcal{K}$. Then, taking the complex conjugate on both sides of the above equation, we have 
\begin{equation}
    h(-k)^T=U_a^* h(k) U_a^T.
\end{equation}
Then note that for the determinant, we have $\det h(k)=\det U_a^* h(k) U_a^T=\det h(-k)^T=\det h(-k)$. We can see that the winding number of $h(k)$ obeys
\begin{align}
    W&=\frac{1}{2\pi\textrm{i}}\int_{BZ} dk\frac{d}{dk}\log \det h(k)\nonumber\\
    &=\frac{1}{2\pi\textrm{i}}\int_{BZ} dk\frac{d}{dk}\log \det h(-k)\nonumber\\
    &=-\frac{1}{2\pi\textrm{i}}\int_{BZ} dk\frac{d}{dk}\log \det h(k)=-W,
\end{align}
and thus, $W=0$.

Next, consider the constraint due to TRS,
\begin{equation}
    h(-k)=\mathcal{T}_b h(k) \mathcal{T}_b^{-1}.
\end{equation}
Using $\mathcal{T}_b=U_b \mathcal{K}$, we arrive at 
\begin{equation}
    h(-k)^*=U_b^* h(k) U_b^T. 
\end{equation}
Then, the determinant of $h(k)$ obeys $\det h(k)=\det h(-k)^*$. We can see that the winding number obeys
\begin{align}
    W&=\frac{1}{2\pi\textrm{i}}\int_{BZ} dk\frac{d}{dk}\log \det h(-k)^*\nonumber\\
    &=-\frac{1}{2\pi\textrm{i}}\int_{BZ} dk\frac{d}{dk}\log \det h(k)^*=W^*,
\end{align}
where in the last step, we used the fact that the winding number is a real number. Therefore, $W$ need not vanish under TRS~\eqref{eq:regular TRS}.

\section{ $\mathbb{Z}_2$ quantization of the Berry phase under TRS} \label{Appendix:Z2 invariant}
In this appendix, we prove that the $\mathbb{Z}_2$ invariant in Table \ref{tab:TopologicalClassification} is the Berry phase quantized to $0$ or $\pi$. 
The constraints of TRS for a generic Hamiltonian are discussed in Sec. \ref{Sec: complex Chern insulators} and Appendix \ref{sec:ConstructionOfClassification}. Now consider a 1D NH Bloch Hamiltonian $h(k)$. As discussed in Sec. \ref{Sec: complex Chern insulators}, TRS forces the eigenvalues of a Hamiltonian $h(k)$ to come in $\{\epsilon_n (-k),\epsilon_n^* (k)\}$ pairs. In the presence of an imaginary line gap, this allows us to label bands of $h(k)$ in pairs $\{\Tilde{n},-\Tilde{n}\}$ for bands above and below the imaginary line gap, respectively. 

Let us now consider the biorthogonal Wilson line defined by 
\begin{align}
    \mathcal{W}_{+,k_f \leftarrow k_i}^{\Tilde{n}}&=G_{k_f-\Delta}^{\Tilde{n}} G_{k_f-2\Delta}^{\Tilde{n}}\cdots\nonumber \\
    &\cdots G_{k_i+\Delta}^{\Tilde{n}} G_{k_i}^{\Tilde{n}},
    \label{eq:WilsonLine}
\end{align}
where the biorthogonal Wilson line element is defined as $[G_k^{\Tilde{n}}]^{mn}=\brakets{u_{k+\Delta}^m}{v_k^n}$. The superscript ${\Tilde{n}}$ labels the band or group of bands over which the Wilson line is calculated, so that $m,n\in \Tilde{n}$. $\Delta$ is the spacing between adjacent Wilson line elements in $k$ space. The sign of $\Delta$ determines the direction in which the Wilson line is calculated. In this appendix, we choose $\Delta>0$, which corresponds to the subscript $+$ on the left-hand side of \eqref{eq:WilsonLine}. 

Let us now consider the case in which the Wilson line traverses the entire BZ, i.e., $k_f=k_i+2\pi$; this constitutes the biorthogonal Wilson loop
\begin{align}
    \mathcal{W}_{+,k}^{\Tilde{n}}=G_{k-\Delta}^{\Tilde{n}} G_{k-2\Delta}^{\Tilde{n}}\cdots \nonumber\\\
        \cdots G_{k+\Delta}^{\Tilde{n}} G_k^{\Tilde{n}},
        \label{eq:NH Wilson loop}
\end{align}
which is gauge-invariant. Furthermore, the eigenvalues of the biorthogonal Wilson loop are independent of the starting point $k$ \cite{PhysRevB.84.075119}. Since we are interested only in the Wilson loop spectrum, we will drop the subscript $k$ in the Wilson loop. Wilson loops calculated by advancing $k$ in opposite directions obey 
\begin{equation}
    \mathcal{W}_{+}^{\Tilde{n}}=[\mathcal{W}_{-}^{\Tilde{n}}]^{-1}. 
    \label{eq:Wilson loop along opposite direction}
\end{equation}
In the Hermitian case, the Wilson loop is unitary, and thus its eigenvalues take the form $\exp(\textrm{i})$, where $\gamma \in \mathbb{R}$ is the \emph{Berry phase}. In the NH case, the Wilson loop is no longer unitary, leading to complex values of $\gamma$. However, in this appendix, we will only consider real $\gamma$ to simplify our argument. Under that consideration, Eq. \eqref{eq:Wilson loop along opposite direction} becomes
\begin{equation}
    \mathcal{W}_{+}^{\Tilde{n}}=[\mathcal{W}_{-}^{\Tilde{n}}]^\dag.
    \label{eq:Wilson loop along opposite direction 2}
\end{equation}
When $\gamma$ is real, the order in which we choose the biorthogonal basis in the definition of the Wilson loop will not affect the result Berry phases, i.e., using $\brakets{u_{k+\Delta}^m}{v_k^n}$ or $\brakets{v_{k+\Delta}^m}{u_k^n}$ for the Wilson line elements will result in the same value of $\gamma$ for the eigenvalues of the Wilson loop.

Now we insert $\mathcal{T}^2=1$ into the Wilson line elements $G_k^{\Tilde{n}}$ and apply Eq. \eqref{eq:Time reversal 1}
\begin{align}
    G_k^{\Tilde{n}}&=\bra{u_{k+\Delta}^{m*}}\mathcal{T}^2 \ket{v_k^{n*}}\nonumber\\
    &= \sum_{-n,-m\in -\Tilde{n}}[V_{k+\Delta}^{-m,m}]^T \brakets{u_{-k-\Delta}^{-m*}}{v_{-k}^{-n*}}[V_k^{-n,n}]^* \nonumber\\
    &=\sum_{-n,-m\in -\Tilde{n}}[V_{k+\Delta}^{-m,m}]^T \brakets{u_{-k-\Delta}^{-m*}}{v_{-k}^{-n*}}[V_k^{-n,n}]^*,
\end{align}
where we defined the sewing matrix $V_{\textbf{k}}^{-n,n}=\bra{v_{-\textbf{k}}^{-n}}\mathcal{T} \ket{u_\textbf{k}^n}$. 
We see that $\brakets{u_{-k-\Delta}^{-m}}{v_{-k}^{-n}}$ is a Wilson line element for the bands in $-\Tilde{n}$ and in the opposite direction. This allows us to exploit the property \eqref{eq:Wilson loop along opposite direction 2}. Applying Eq. \eqref{eq:NH Wilson loop} and using the fact that $[V_k^{-n,n}]^\dag V_k^{-n,n}=1$, we get
\begin{equation}
    \mathcal{W}_{+}^{\Tilde{n}}=V_k^\dag \mathcal{W}_{+}^{-\Tilde{n}} V_k.
\end{equation}
This implies that the Berry phases $\gamma_{\Tilde{n}}$ for bands in $\Tilde{n}$ obey $\{e^{\textrm{i} \gamma_{\Tilde{n}}}\}=\{e^{\textrm{i} \gamma_{-\Tilde{n}}}\}$. Consider now a system with only two bands. The two Berry phases obey
\begin{equation}
    \gamma_{\Tilde{n}}=\gamma_{-\Tilde{n}}.
    \label{eq:constraint1}
\end{equation}
For such systems, there is the additional constraint that the Wilson loop for the combined bands $\Tilde{n}$ and $-\Tilde{n}$ is trivial, such that
\begin{equation}
\gamma_{\Tilde{n}}+\gamma_{-\Tilde{n}}=0 \mod{2\pi}. 
\label{eq:constraint2}
\end{equation}
Equations~\eqref{eq:constraint1} and \eqref{eq:constraint2} lead to two possible values for the Berry phase, $\gamma_{\Tilde{n}}=0$ or $\pi$. This is the $\mathbb{Z}_2$ invariant in 38-fold classification table for 1D in class AI or Table \ref{tab:TopologicalClassification} in this paper. In the case of systems with multiple bands, it is straightforward to generalize our findings and show that the quantized index is the polarization, $p=\frac{1}{2\pi}\log \det (\mathcal{W})$.

\section{Details on the deformation of complex Chern insulators into HOSE phases} \label{Appendix:DeformationImGap}

In this appendix, we consider a model that connects the HOSE phase [Eq. \eqref{eq:HOSE}] with the imaginary-line-gap Chern insulator [Eq. \eqref{eq:ComplexChernInsulatorImGap}]. 
Consider a lattice with the two-band Bloch Hamiltonian
\begin{align}
    h^\mathrm{Im}_\mathrm{def}({\bf k})=&-\textrm{i}\sigma_z \sin k_x+\textrm{i} \sigma_x \sin k_y \nonumber\\
    &+\textrm{i} \sigma_y (\cos k_y+t \cos k_x+m)\nonumber\\
    &+\sigma_0 \cos k_x (1-g).
    \label{eq:DeformationImGap}
\end{align}
This model obeys TRS \eqref{eq:regular TRS} with $\mathcal{T}=\mathcal{K}$. Just as in the deformation of the real-line-gap Chern insulator to a HOSE phase described in Section \ref{sec:DeformationComplexChern}, we set $m=0.5$ throughout the entire deformation. The main difference between the deformation in the main text and this one is that this one is in class AI. We first set the parameters to $(g,t)=(1,1)$, which puts Hamiltonian~\eqref{eq:DeformationImGap} in the imaginary-line-gap Chern insulator phase [Fig.~\ref{fig:ComplexChernInsulators}(b)]. We then deform Hamiltonian~\eqref{eq:DeformationImGap} according to $(g,t): (1,1)-\theta (1,1)$ for $\theta \in [0,1]$. 

\begin{figure}
    \centering
    \includegraphics[width=\columnwidth]{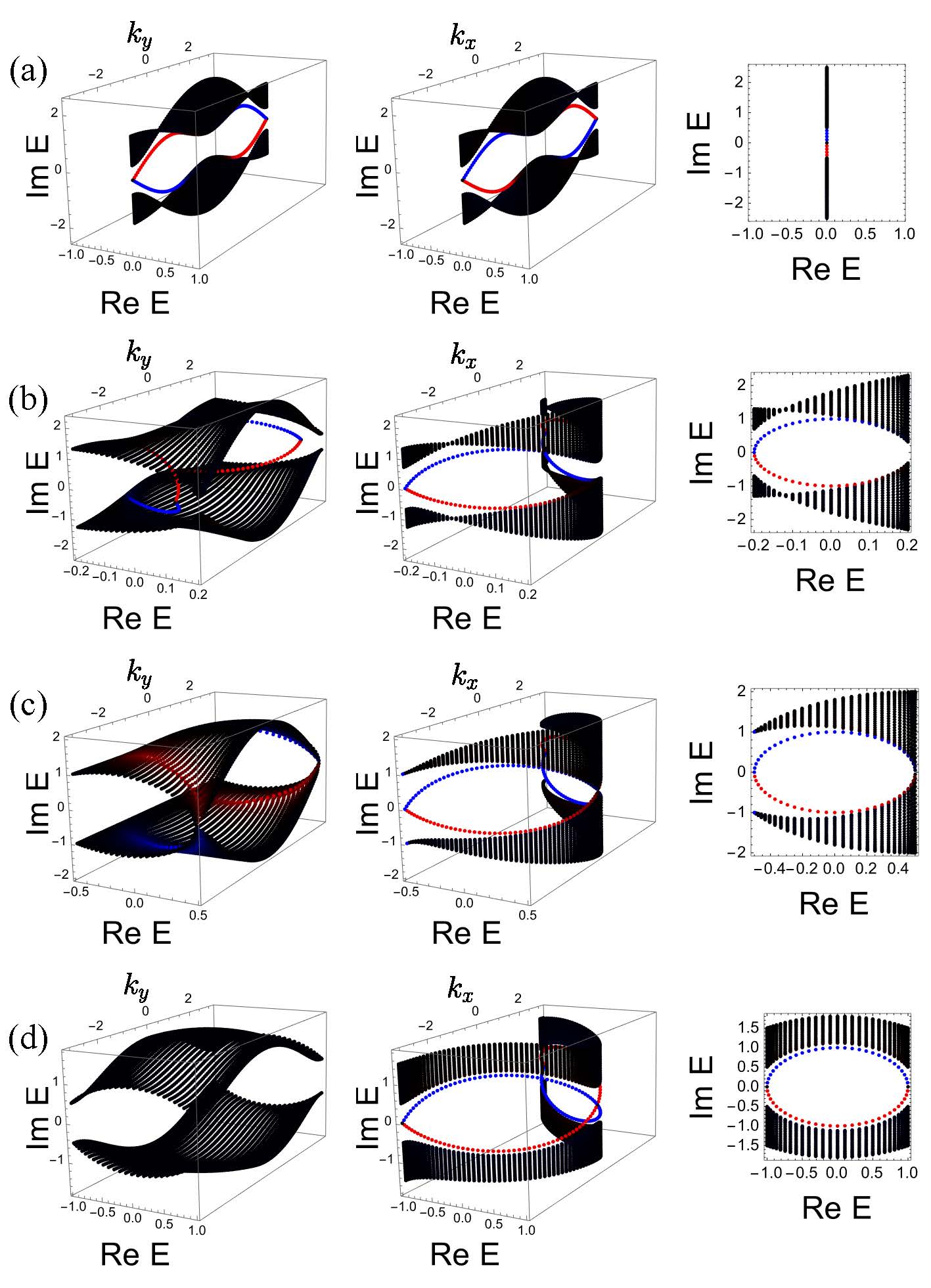}
    \caption{Deforming an imaginary-line-gap complex Chern insulator into a HOSE phase in the Bloch Hamiltonian \eqref{eq:DeformationImGap}. (a)--(d) correspond to $\theta=0,0.2,0.5,1$, respectively. (a,b) An imaginary-line-gap complex Chern insulator. (c) Bulk phase transition between the imaginary-line-gap complex Chern insulator phase and the HOSE phase. (d) HOSE phase. Left panels: OBC along $x$ and PBC along $y$; middle panels: OBC along $y$ and PBC along $x$; right panels: band projections of the plots on the second column on the complex energy plane. Black represents bulk states, while blue and red denote states localized at opposite edges.}
    \label{fig:DeformationImGap}
\end{figure}

We first continuously evolve $\theta$ from $0$ to $0.2$. The spectrum at $\theta=0.2$ is shown in Fig.~\ref{fig:DeformationImGap}(b). Next, we evolve $\theta$ from $0.2$ to $0.5$. A phase transition occurs at $\theta=0.5$ as shown in Fig.~\ref{fig:DeformationImGap}(c). Finally, we vary $\theta$ from $0.5$ to $1$. As shown in Fig.~\ref{fig:DeformationImGap}(d), this model is in a HOSE phase.

\begin{figure*}
    \centering
    \includegraphics[width=\textwidth]{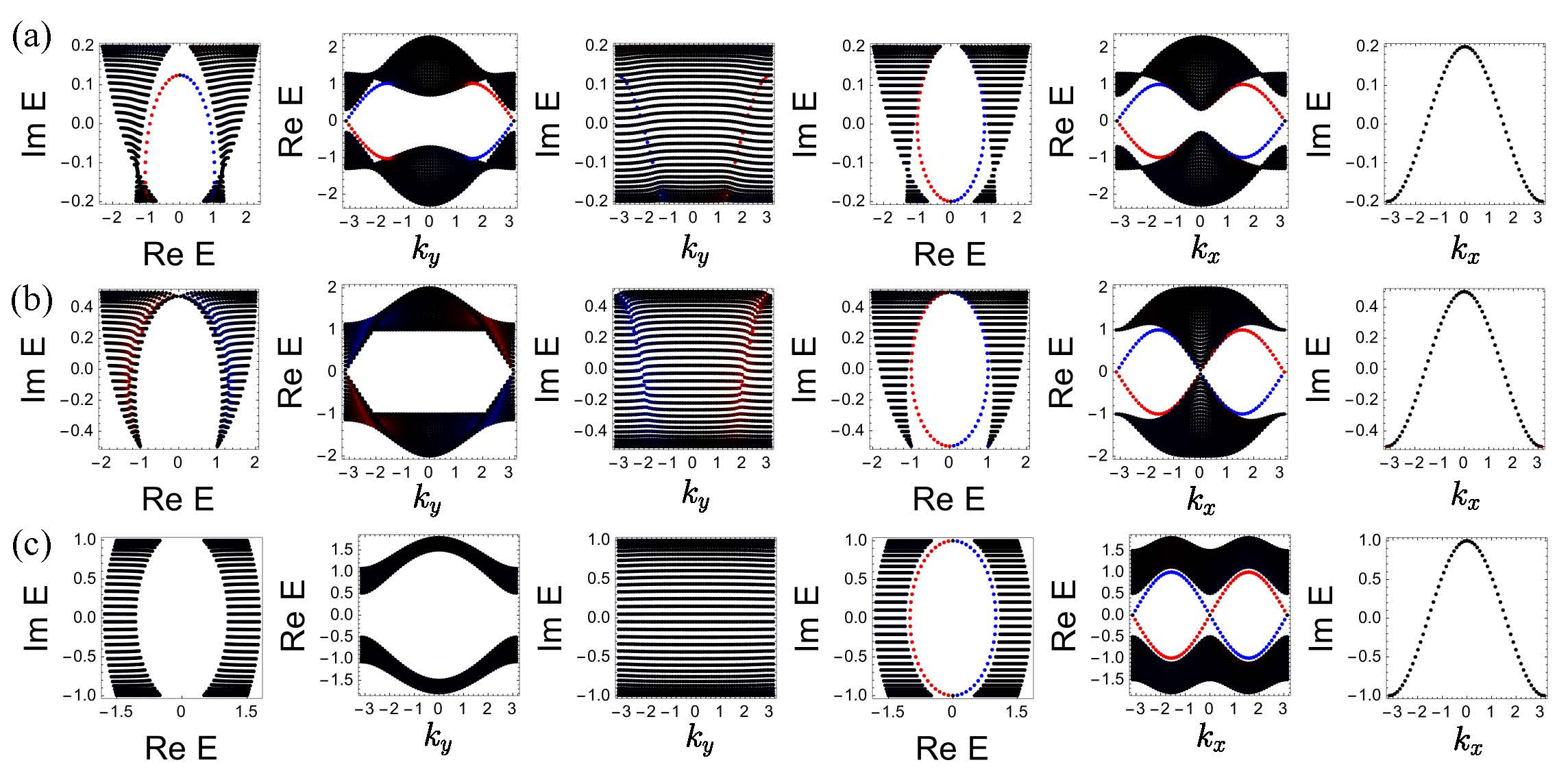}
    \caption{Spectrum of model with Bloch Hamiltonian~\eqref{eq:DeformationReGap}. For all plots, left (right) three panels are spectra for OBC only along $x$ ($y$). The deformation path is $(g,t,m)=(1,1,0.5)-\theta(1,1,0)$. (a) Chern phase at $\theta=0.2$. (b) Phase transition at $\theta=0.5$. (c) HOSE phase at $\theta=1$.}
    \label{fig:AppendixDeformationReGap}
\end{figure*}

\begin{figure*}
    \centering
    \includegraphics[width=\textwidth]{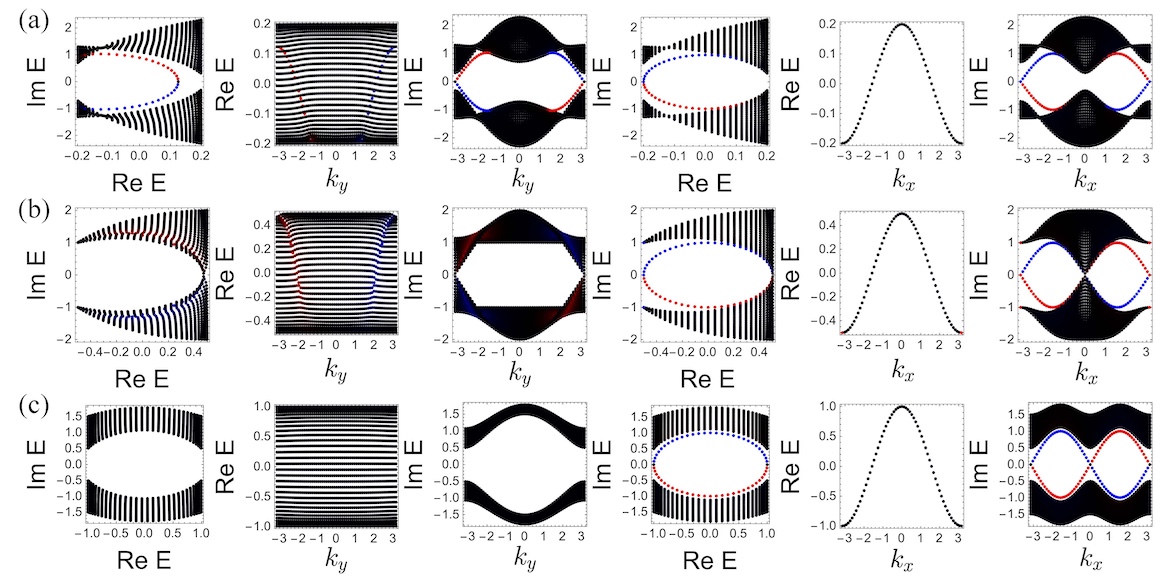}
    \caption{Spectrum of model with Bloch Hamiltonian~\eqref{eq:DeformationImGap}. For all plots, left(right) three panels are spectra for OBC only along $x$ ($y$). The deformation path is $(g,t,m)=(1,1,0.5)-\theta(1,1,0)$. (a) Imaginary-line-gap complex Chern phase at $\theta=0.2$. (b) Phase transition at $\theta=0.5$. (c) HOSE phase at $\theta=1$.}
    \label{fig:AppendixDeformationImGap}
\end{figure*}

Figures~\ref{fig:AppendixDeformationReGap} and \ref{fig:AppendixDeformationImGap} show the spectrum for some points during the deformations in~\eqref{eq:DeformationReGap} and~\eqref{eq:DeformationImGap} that complement those shown in Figs.~\ref{fig:DeformationReGap} and \ref{fig:DeformationImGap}. As we can confirm from (b) in Figs.~\ref{fig:AppendixDeformationReGap} and \ref{fig:AppendixDeformationImGap}, the deformation is not smooth since line gaps close, causing a phase transition.


\section{Construction of the topological classification of $C_n$-symmetric NH Hamiltonians in class AI}\label{sec:ConstructionOfClassification}
\begin{figure}
    \centering
    \includegraphics[width=.9\columnwidth]{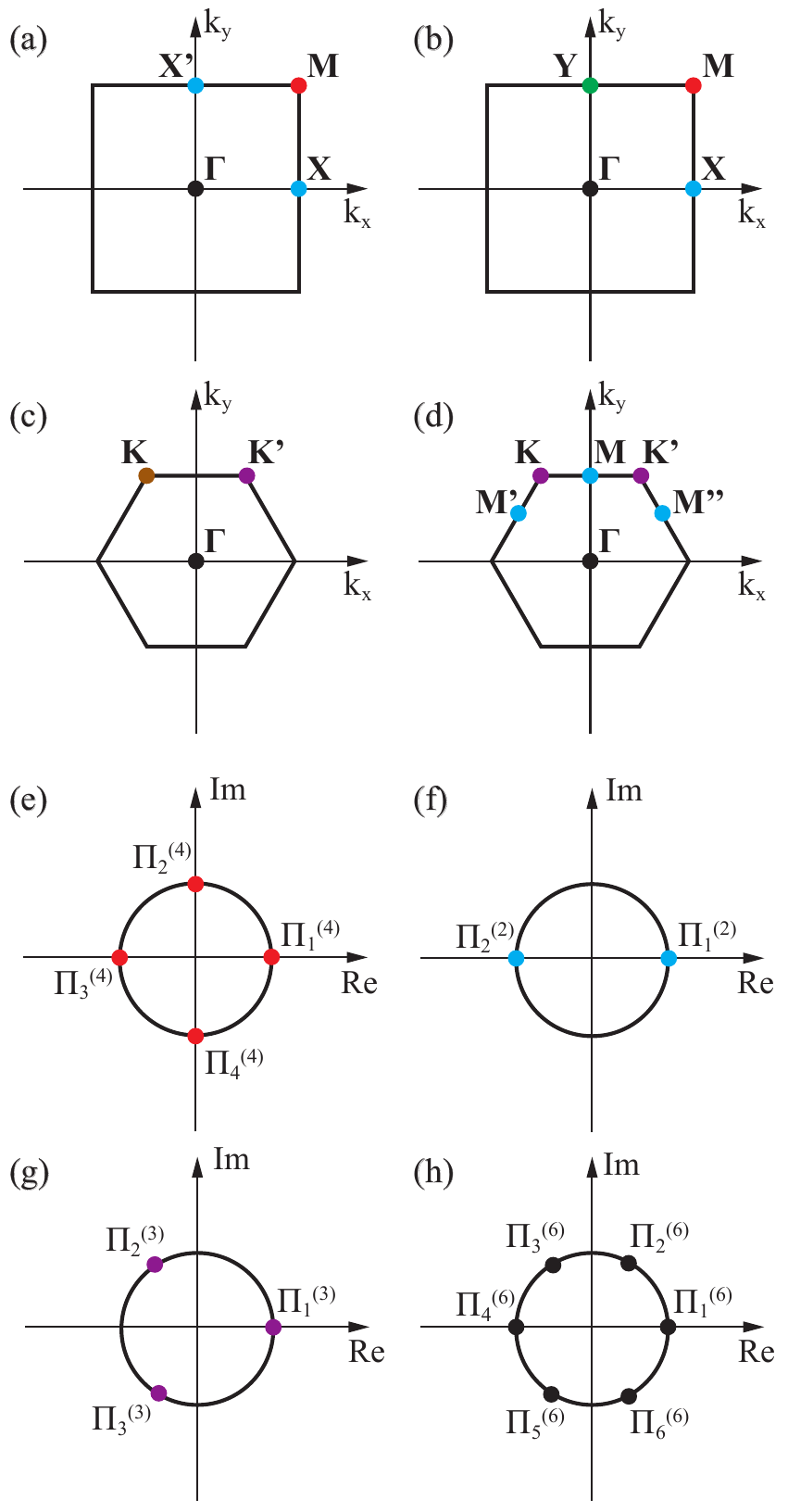}
    \caption{(a)-(d) HSPs in the BZ of $C_{4,2,3,6}$-symmetric lattices, respectively. Colored dots are the HSPs defined in Eq. \eqref{eq:HSPs def}. (e)-(h) Eigenvalues of the rotation operators  $\hat{r}_4$, $\hat{r}_2$, $\hat{r}_3$, and $\hat{r}_6$, respectively. In (e)-(h), we consider only operators obeying $[\hat{r}_n]^n=1$.}
    \label{fig:Cn symmetry}
\end{figure}
In this appendix, we build the topological indices of Eq.~\eqref{eq:IndicesComplexBands} that classify the energy bands of $C_n$-symmetric NH lattices in class AI of the 38-fold way. Because of non-Hermiticity, the classification is, in general, different from the one obtained for Hermitian systems. However, for ``real'' energy bands, additional constraints result in indices~\eqref{eq:IndicesRealBands}, which coincide with the indices of Hermitian energy bands~\cite{benalcazar_quantization_2019}.
We first discuss the implications of TRS and $C_n$ rotation symmetry on the energy bands and build the symmetry indicator invariants. Then, we discuss the constraints these two symmetries impose on these invariants. These two steps then allow the construction of the $\chi^{(n)}$ indices in Eq.~\eqref{eq:IndicesComplexBands} and \eqref{eq:IndicesRealBands}.  

\subsection{Time reversal symmetry} 
In Sec. \ref{Sec: complex Chern insulators}, we saw that TRS enforces the relation
\begin{equation}
    h(-\textbf{k})\mathcal{T} \ket{u_{\textbf{k}}^l}=\mathcal{T}h(\textbf{k})\ket{u_{\textbf{k}}^l}=\epsilon_l^*(\textbf{k})\mathcal{T}\ket{u_{\textbf{k}}^l}.
    \label{eq:D1}
\end{equation}
Hence, $\mathcal{T}\ket{u_{\textbf{k}}^l}$ is an eigenstate of $h(-\textbf{k})$ with energy $\epsilon_l^*(\textbf{k})$. 
To prove the constraint of TRS on the energy eigenvalues more rigorously, we project the state at $l \in \Tilde{l}$ into the space spanned by states at band $-\Tilde{l}$.
\begin{align}
   \mathcal{T}\ket{u_\textbf{k}^l}=\sum_{-l\in-\Tilde{l}}\ket{u_{-\textbf{k}}^{-l}}\bra{v_{-\textbf{k}}^{-l}}\mathcal{T} \ket{u_\textbf{k}^l}=\sum_{-l\in-\Tilde{l}}V_{\textbf{k}}^{-l,l}\ket{u_{-\textbf{k}}^{-l}},
   \label{eq:Appendix:TRS}
\end{align}
where we have defined sewing matrix $V_{\textbf{k}}^{-l,l}=\bra{v_{-\textbf{k}}^{-l}}\mathcal{T} \ket{u_\textbf{k}^l}$. 

From Eq.~\eqref{eq:D1}, it follows that
\begin{equation}
     h(-\textbf{k})\mathcal{T} \ket{u_{\textbf{k}}^l}=\epsilon_l^*(\textbf{k})\mathcal{T} \ket{u_{\textbf{k}}^l}=\epsilon_l^*(\textbf{k})\sum_{-\Tilde{l}}V_{\textbf{k}}^{-l,l}\ket{u_{-\textbf{k}}^{-l}}.
\end{equation}
On the other hand, we have
\begin{align}
    h(-\textbf{k})\mathcal{T} \ket{u_{\textbf{k}}^l}&=h(-\textbf{k})\sum_{-\Tilde{l}}V_{\textbf{k}}^{-l,l}\ket{u_{-\textbf{k}}^{-l}}\nonumber\\
    &=\sum_{-\Tilde{l}}\epsilon_{-l}(-\textbf{k})V_{\textbf{k}}^{-l,l}\ket{u_{-\textbf{k}}^{-l}}.
\end{align}
Therefore, by subtracting these two equations, we have
\begin{align}
    \sum_{-\Tilde{l}}V_{\textbf{k}}^{-l,l}\ket{u_{-\textbf{k}}^{-l}}(\epsilon_l^*(\textbf{k})-\epsilon_{-l}(-\textbf{k}))=0
\end{align}
for every $l \in \Tilde{l}$. Applying $\bra{v^{-l}_{-\textbf{k}}}$ to the above expression, we get
\begin{equation}
    V_{\textbf{k}}^{-l,l}(\epsilon_l^*(\textbf{k})-\epsilon_{-l}(-\textbf{k}))=0
\end{equation}
for every $l \in \Tilde{l}$ and $-l \in -\Tilde{l}$, which implies that the sewing matrix has elements $V_{\textbf{k}}^{-l,l} \neq 0$ only for bands obeying $\epsilon_l^*(\textbf{k})=\epsilon_{-l}(-\textbf{k})$.

\subsection{Rotation symmetry}
Rotation symmetry is expressed as
\begin{equation}
    \hat{r}_n h(\textbf{k})\hat{r}_n^\dag=h(R_n\textbf{k}),
    \label{eq:rotation symmetry appendix}
\end{equation}
where $\hat{r}_n$ is the $n$-fold rotation operator, which obeys $[\hat{r}_n]^n=\pm1$ and $R_n$ is the $n$-fold rotation matrix acting on the crystal momentum $\textbf{k}$. 
Let Eq. \eqref{eq:rotation symmetry appendix} act on the energy eigenstate of band $l\in \Tilde{l}$. We have
\begin{equation}
    h(R_n\textbf{k})\hat{r}_n\ket{u_\textbf{k}^l}=\hat{r}_n h(\textbf{k})\ket{u_\textbf{k}^l}=\epsilon_l(\textbf{k})\hat{r}_n\ket{u_\textbf{k}^l}.
    \label{eq:rotation connection eq}
\end{equation}
Thus $\hat{r}_n\ket{u_\textbf{k}^l}$ is an eigenstate of $h(R_n\textbf{k})$ with eigenvalue $\epsilon_l(\textbf{k})$. We can make the expansion
\begin{equation}
\hat{r}_n\ket{u_\textbf{k}^l}=\sum_{q \in \Tilde{l}}\ket{u_{R_n\textbf{k}}^q}B_\textbf{k}^{ql}
\end{equation}
where the sewing matrix of the rotation operator is defined as
\begin{equation}
B_\textbf{k}^{ql}=\bra{v_{R_n\textbf{k}}^q}\hat{r}_n\ket{u_\textbf{k}^l}.
    \label{eq:sewing B}
\end{equation}
High-symemtry points (HSPs) ${\bf \Pi}_m$ remain invariant under the-little group $C_m$ rotation modulo a reciprocal lattice vector ${\bf G}$ (for $m \leq n$), i.e., they obey
\begin{equation}
    R_m{\bf \Pi}_m={\bf \Pi}_m \quad (\text{mod } {\bf G}). 
    \label{eq:HSPs def0}
\end{equation}
The HSPs in the BZ of $C_n$-symmetric lattices are
\begin{align}
   &C_2:\quad {\bf X}=\{\pi,0\}, {\bf Y}=\{0,\pi\}, {\bf M}=\{\pi,\pi\},\nonumber\\
   &C_4:\quad {\bf X}=\{\pi,0\}, {\bf X'}=\{0,\pi\}, {\bf M}=\{\pi,\pi\},\nonumber\\
   &C_3:\quad {\bf K}=\{-\frac{2\pi}{3},\frac{2\pi}{\sqrt{3}}\},{\bf K}'=\{\frac{2\pi}{3},\frac{2\pi}{\sqrt{3}}\},\nonumber\\
   &C_6:\quad {\bf K}=\{-\frac{2\pi}{3},\frac{2\pi}{\sqrt{3}}\},{\bf K}'=\{\frac{2\pi}{3},\frac{2\pi}{\sqrt{3}}\},\nonumber\\
   &{\bf M}=\{0,\frac{2\pi}{\sqrt{3}}\},{\bf M}'=\{-\pi,\frac{\pi}{\sqrt{3}}\},{\bf M}''=\{\pi,\frac{\pi}{\sqrt{3}}\},
    \label{eq:HSPs def}
\end{align}
and ${\bf \Gamma}=\{0,0\}$ for all cases. These HSPs are indicated in Figs.~\ref{fig:Cn symmetry}(a)--\ref{fig:Cn symmetry}(d). In $C_2$-symmetric lattices, ${\bf X}$, ${\bf Y}$, ${\bf M}$ are invariant under $C_2$ rotations; in $C_4$-symmetric lattices, ${\bf M}$ is invariant under $C_4$ while ${\bf X}$ and ${\bf X}'$ are invariant under $C_2$; in $C_3$-symmetric lattices, ${\bf K}$ and ${\bf K}'$ are invariant under $C_3$; and in  $C_6$-symmetric lattices, ${\bf K}$ and ${\bf K}'$ are invariant under $C_3$ while ${\bf M}$, ${\bf M}'$, and ${\bf M}''$ are invariant under $C_2$. Finally, ${\bf \Gamma}$ is invariant under the full group $C_n$ rotation for all $C_n$ symmetric lattices.

From Eq.~\eqref{eq:HSPs def0} and~\eqref{eq:rotation symmetry appendix} it follows that $[h({\bf \Pi}_m),\hat{r}_m]=0$. Thus, at HSPs, we also have
\begin{equation}
    \hat{r}_m\ket{u_{{\bf \Pi}_m}^l}=r_{{\bf \Pi}_m}^l\ket{u_{{\bf \Pi}_m}^l},
\end{equation}
where $r_{{\bf \Pi}_m}^l$ is the rotation eigenvalue associated with energy band $\Tilde{l}$ at HSP ${\bf \Pi}_m$, which can take the values
\begin{equation}
    {\Pi}_p^{(m)}=
    \begin{cases}
        e^{2\pi \textrm{i}(p-1)/m}, & \text{for }[\hat{r}_n]^n=1\\
        e^{2\pi \textrm{i}(p-1/2)/m}, & \text{for }[\hat{r}_n]^n=-1
    \end{cases}
\end{equation}
for $p=1,2,\dots m$.
We now define the symmetry indicator invariants
\begin{equation}
    [{\Pi}_p^{(m)}]_{\Tilde{l}}=\#_{\Tilde{l}}{\Pi}_p^{(m)}-\#_{\Tilde l}\Gamma_p^{(m)}, \label{eq:app_RotationInvariants}
\end{equation}
where $\#_{\Tilde{l}}{\Pi}_p^{(m)}$ is the number of energy bands in the band group $\Tilde{l}$ with eigenvalue ${\Pi}_p^{(m)}$. Note that if there is an equal number of bands of a given rotation representation of $\hat{r}_m$ at both ${\bf \Pi}_m$  and ${\bf \Gamma}=(0,0)$, the symmetry indicator invariants are zero. Hence, these symmetry indicator invariants signal an imbalance in the number of representations across a generic HSP and those at ${\bf \Gamma}$ at energy bands $\Tilde{l}$. Not all these invariants are independent. In a $C_4$ symmetric crystal, rotation symmetry forces the representation at $\bf X$ and $\bf X'$ to be equal [Fig.~\ref{fig:Cn symmetry}(a)]. Similarly, $C_6$ symmetry forces equal representations at $\bf M$, $\bf M'$, and $\bf M''$, as well as at $\bf K$ and $\bf K'$ [Fig.~\ref{fig:Cn symmetry}(d)]. We will demonstrate this in the following sections. 

\subsection{Constraints due to rotation}\label{Appendix:rotation constraint}
Consider a crystal with $C_{n}$ symmetry with operator $\hat{r}_n$. $C_n$ symmetry relates some of the HSPs~\eqref{eq:HSPs def} that are invariant under little group $C_m$, where $m<n$. We are interested in the eigenvalues of the $C_m$ rotation operator at ${\bf \Pi}_m$ and $R_n{\bf \Pi}_m$. We now demonstrate that the symmetry indicator invariants for $C_{m}$ of a band group $\Tilde{l}$ at $m$-fold HSPs are identical due to $C_{n}$.  

Since $R_n {\bf \Pi}_m$ is invariant under $R_m$, we have
\begin{equation}
    \hat{r}_m\ket{u_{R_n{\bf \Pi}_m}^l}= r_{R_n{\bf \Pi}_m}^l\ket{u_{R_n{\bf \Pi}_m}^l},
\end{equation}
for band $l\in\Tilde{l}$. 
Since $R_n{\bf \Pi}_m$ and ${\bf \Pi}_m$ are related by $C_{n}$ symmetry, we can make the expansion
\begin{equation}
    \hat{r}_n\ket{u_{{\bf \Pi}_m}^l}=\sum_{q\in\Tilde{l}}\ket{u_{R_n{\bf \Pi}_m}^q}B_{{\bf \Pi}_m}^{ql},
\end{equation}
where $B_{{\bf \Pi}_{m}}^{ql}=\bra{v_{R_{n}{\bf \Pi}_m}^q}\hat{r}_n\ket{u_{{\bf \Pi}_m}^l}$ is the sewing matrix of rotation at HSP ${\bf \Pi}_m$. By applying $\hat{r}_m$ to the above expression and using the fact that $[\hat{r}_m,\hat{r}_n]=0$, we have
\begin{equation}
    (r_{R_n{\bf \Pi}_m}^q-r_{{\bf \Pi}_m}^l)B_{{\bf \Pi}_m}^{ql}=0
\end{equation}
for all $q,l\in \Tilde{l}$. Thus the rotation eigenvalues of the little group at $R_n{\bf \Pi}_m$ and ${\bf \Pi}_m$ are equal at any given band groups $\Tilde{l}$,
\begin{equation}
\boxed{
    \{r_{R_n{\bf \Pi}_m}^l\}_{\Tilde{l}}\overset{C_n}{=}\{r_{{\bf \Pi}_m}^l\}_{\Tilde{l}}
    }
    \label{eq:Cn symmetry constraint}
\end{equation}
More explicitly,
\begin{align}
    &\{r_{\textbf{X}}^l\}\overset{C_4}{=}\{r_{\textbf{X}'}^l\},\nonumber\\
    &\{r_{\textbf{K}}^l\}\overset{C_6}{=}\{r_{\textbf{K}'}^l\},\nonumber\\
    &\{r_{\textbf{M}}^l\}\overset{C_6}{=}\{r_{\textbf{M}'}^l\}\overset{C_6}{=}\{r_{\textbf{M}''}^l\},
\end{align}
This implies that the invariants \eqref{eq:app_RotationInvariants} obey
\begin{align}
    &[X_p^{(2)}]\overset{C_4}{=}[X_p^{'(2)}],\nonumber\\ 
    &[K_p^{(3)}]\overset{C_6}{=}[K_p^{'(3)}],\nonumber\\
    &[M_p^{(2)}]\overset{C_6}{=}[M_p^{'(2)}]\overset{C_6}{=}[M_p^{''(2)}],
\end{align}
in the same band group $\Tilde{l}$. This conclusion applies for both $[\hat{r}_n]^n=\pm1$ cases.
 
\subsection{Constraints due to TRS}
TRS will add another constraint to the rotation invariants \eqref{eq:app_RotationInvariants} $[{\Pi}_p^{(m)}]$. The TRS operator and rotation operator in general commute
\begin{equation}
    [\mathcal{T},\hat{r}_m]=0.
\end{equation}
Thus, we have
\begin{align}
    \mathcal{T} (\hat{r}_m\ket{u_\textbf{k}^l})&=\mathcal{T}\sum_{q\in \Tilde{l}}\ket{u_{R_m\textbf{k}}^{q}}B_\textbf{k}^{q,l}\nonumber\\
    &=\sum_{-q\in -\Tilde{l},q\in \Tilde{l}}\ket{u_{-R_m\textbf{k}}^{-q}}V_{R_m\mathbf{k}}^{-q,q}B_\mathbf{k}^{q,l*}.
\end{align}
Here, $B_\textbf{k}^{q,l}=\bra{v_{R_m\textbf{k}}^q}\hat{r}_m\ket{u_\textbf{k}^l}$ is the sewing matrix of $\hat{r}_m$ from the little group $C_m$.
On the other hand, we have
\begin{align}
    \hat{r}_m(\mathcal{T}\ket{u_\textbf{k}^l})&=\hat{r}_m\sum_{-l\in-\Tilde{l}}\ket{u_{-\textbf{k}}^{-l}}V_{\textbf{k}}^{-l,l}\nonumber\\
    &=\sum_{-q,-l\in-\Tilde{l}}\ket{u_{-R_m\textbf{k}}^{-q}}B_\mathbf{-k}^{-q,-l}V_{\mathbf{k}}^{-l,l}.
\end{align}
Therefore, by subtracting these two equations and acting $\bra{v_{-R_m\textbf{k}}^{-q}}$ on the left, we have
\begin{equation}
    \sum_{q\in \Tilde{l},-l\in -\Tilde{l}}V_{R_m\mathbf{k}}^{-q,q}B_\mathbf{k}^{q,l*}-B_\mathbf{-k}^{-q,-l}V_{\mathbf{k}}^{-l,l}=0.
\end{equation}
At HSPs ${\bf \Pi}_m$, we choose the gauge in which $B_{{\bf \Pi}_m}^{l,q}=r_{{\bf \Pi}_m}^l\delta_{l,q}$ is diagonal, we have
\begin{equation}
    V_{{\bf \Pi}_m}^{-q,l}(r_{{\bf \Pi}_m}^{l*}-r_{-{\bf \Pi}_m}^{-l})=0.
\end{equation}
Hence, TRS imposes the constraint

\begin{equation}
    \boxed{\{r_{{\bf \Pi}_m}^l\}_{\Tilde{l}}\overset{TRS}{=}\{r_{-{\bf \Pi}_m}^{-l*}\}_{-\Tilde{l}}}
    \label{eq:TRS constraint}
\end{equation}
For the symmetry indicator invariants, this constraint implies the following relations:

For $[\hat{r}_n]^n=1$,
\begin{align}
    &[M_2^{(4)}]_{\pm\Tilde{l}}\overset{C_4}{=}[M_4^{(4)}]_{\mp \Tilde{l}},\nonumber\\ 
    &[K_1^{(3)}]_{\pm\Tilde{l}}\overset{C_{3.6}}{=}[K_1^{'(3)}]_{\mp \Tilde{l}},\nonumber\\
    &[K_2^{(3)}]_{\pm\Tilde{l}}\overset{C_{3,6}}{=}[K_3^{'(3)}]_{\mp \Tilde{l}},\nonumber\\
    &[K_3^{(3)}]_{\pm\Tilde{l}}\overset{C_{3,6}}{=}[K_2^{'(3)}]_{\mp \Tilde{l}}.
\end{align}
For $[\hat{r}_n]^n=-1$,
\begin{align}
    &[M_1^{(4)}]_{\pm\Tilde{l}}\overset{C_4}{=}[M_4^{(4)}]_{\mp \Tilde{l}},\nonumber\\ 
    &[M_2^{(4)}]_{\pm\Tilde{l}}\overset{C_4}{=}[M_3^{(4)}]_{\mp \Tilde{l}},\nonumber\\ 
    &[X_1^{(2)}]_{\pm\Tilde{l}}\overset{C_{2,4}}{=}[X_2^{(2)}]_{\mp\Tilde{l}},\nonumber\\
    &[M_1^{(2)}]_{\pm\Tilde{l}}\overset{C_{6}}{=}[M_2^{(2)}]_{\mp\Tilde{l}},\nonumber\\
    &[K_1^{(3)}]_{\pm\Tilde{l}}\overset{C_{3,6}}{=}[K_3^{'(3)}]_{\mp \Tilde{l}},\nonumber\\
    &[K_3^{(3)}]_{\pm\Tilde{l}}\overset{C_{3,6}}{=}[K_1^{'(3)}]_{\mp \Tilde{l}},\nonumber\\
    &[K_2^{(3)}]_{\pm\Tilde{l}}\overset{C_{3,6}}{=}[K_2^{'(3)}]_{\mp \Tilde{l}}.
\end{align}
For the rest of the symmetry indicator invariants, i.e., those corresponding to real-valued rotation eigenvalues, the constraint is
\begin{align}
    [\Pi_p^{(m)}]_{\pm\Tilde{l}}\overset{C_n}{=}[\Pi_p^{(m)}]_{\mp\Tilde{l}}
\end{align}
for both $[\hat{r}_n]^n=\pm1$. 

\subsection{$\chi^{(n)}$ indices for $C_n$-symmetric NH crystals}
Since the number of bands within each band group $\Tilde{l}$ is constant across the BZ, we have the constraint
\begin{equation}
\boxed{
    \sum_p [{\Pi}_p^{(m)}]_{\Tilde{l}}=0.
    }
\end{equation}
The three boxed equations above give the complete set of constraints on the symmetry indicator invariants $\{[{ \Pi}_p^{(m)}]\}$. The only difference between the non-Hermitian and the Hermitian cases lies in Eq.~\eqref{eq:TRS constraint} for ``non-real bands'', i.e., when $\Tilde{l}\neq -\Tilde{l}$, in which case TRS relates energy eigenstates across the imaginary line gap. If $\Tilde{l}=-\Tilde{l}$, i.e., if the energy band groups are real, the classification is the same as in the Hermitian case. 
Based on the above discussion, we can now generalize the $\chi^{(n)}$ indices introduced in Ref. \cite{benalcazar_quantization_2019} to NH systems: 

\subsubsection{Case $\Tilde{l}\neq -\Tilde{l}$}

The full classification is given by
\begin{align}
    \chi^{(2)}&=(C|[X_1^{(2)}],[Y_1^{(2)}],[M_1^{(2)}]),\nonumber\\
   \chi^{(4)} &=(C|[X_1^{(2)}],[M_1^{(4)}],[M_2^{(4)}],[M_3^{(4)}]),\nonumber\\
    \chi^{(3)} &=(C|[K_1^{(3)}],[K_2^{(3)}],[K_1^{'(3)}],[K_2^{'(3)}]),\nonumber\\
    \chi^{(6)}&=(C|[M_1^{(2)}],[K_1^{(3)}],[K_2^{(3)}]),
    \label{eq: Appendix Topological invariant}
\end{align}
where, for a set of non-redundant invariants, it suffices to determine the $\chi^{(n)}$ only for energy band groups $\Tilde{l}$ such that $\text{Im}[\epsilon_{\Tilde{l}}]>0$.

\subsubsection{Case $\Tilde{l}= -\Tilde{l}$}

For real-energy bands, i.e., those for which $\Tilde{l} = -\Tilde{l}$, some of the symmetry indicator invariants are redundant, i.e., they can be obtained from other symmetry indicator invariants at the same HSP. Specifically, while for real bands with $[\hat{r}_n]^n=1$, $[M_3^{(4)}]=-2[M_2^{(4)}]-[M_1^{(4)}]$ in $\chi^{(4)}$, $([K_1^{'(3)}],[K_2^{'(3)}])=([K_1^{(3)}],-[K_1^{(3)}]-[K_2^{(3)}])$ in $\chi^{(3)}$, and $[K_2^{(3)}]=-[K_1^{(3)}]/2$ in $\chi^{(6)}$; for real bands with $[\hat{r}_n]^n=-1$, all twofold rotation symmetry indicator invariants are $0$, $[M_3^{(4)}]=[M_2^{(4)}]$, $[M_2^{(4)}]=[M_1^{(4)}]$ in $\chi^{(4)}$, $([K_1^{'(3)}],[K_2^{'(3)}])=(-[K_1^{(3)}]-[K_2^{(3)}],[K_2^{(3)}])$ in $\chi^{(3)}$, and $[K_2^{(3)}]=-[K_1^{(3)}]/2$ in $\chi^{(6)}$. Dropping redundant indicators, and noting that $C=0$ for these bands, the classification for real-energy bands with $[\hat{r}_n]^n=1$ is reduced to 
\begin{align}
    \chi^{(2)}_\textrm{Real}&=(0 | [X_1^{(2)}],[Y_1^{(2)}],[M_1^{(2)}]),\nonumber\\
    \chi^{(4)}_\textrm{Real}&=(0 | [X_1^{(2)}],[M_1^{(4)}],[M_2^{(4)}]),\nonumber\\
    \chi^{(3)}_\textrm{Real}&=(0 | [K_1^{(3)}],[K_2^{(3)}]),\nonumber\\
    \chi^{(6)}_\textrm{Real}&=(0 | [M_1^{(2)}],[K_1^{(3)}]),
    \label{eq:IndicesRealBands}
\end{align}
which coincides with the classification of Hermitian Hamiltonians~\cite{benalcazar_quantization_2019}. For real-energy bands with $[\hat{r}_n]^n=-1$, the classification is reduced to
\begin{align}
    \chi^{(2)}_\textrm{Real}&=(0|0),\nonumber\\
   \chi^{(4)}_\textrm{Real} &=(0|[M_1^{(4)}]),\nonumber\\
    \chi^{(3)}_\textrm{Real} &=(0|[K_1^{(3)}],[K_2^{(3)}]),\nonumber\\
    \chi^{(6)}_\textrm{Real}&=(0|[K_1^{(3)}]).
\end{align}
The subscript $\textrm{Real}$ indicates these invariants are calculated for energy bands on the real energy line, i.e., those obeying $\Tilde{l}= -\Tilde{l}$.

\section{$\chi^{(2)}$ index of the minimal model for a HOSE phase}\label{Appendix: chi2 index HOSE}
As mentioned in the main text, the crystal with Hamiltonian $h_{\text{HOSE}}(\textbf{k})$ in~\eqref{eq:HOSE} obeys $C_2$ symmetry with operator $\hat{r}_2=\sigma_y$. However, it does not obey TRS \eqref{eq:regular TRS}, and thus it does not belong to class AI. The Hamiltonian $\textrm{i} h_{\text{HOSE}}(\textbf{k})$, on the other hand, obeys TRS~\eqref{eq:regular TRS}, with $\mathcal{T}=\mathcal{K}$. This Hamiltonian is
\begin{align}
    \textrm{i}h_{\text{HOSE}}(\textbf{k})=&\cos k_x\sigma_0+\textrm{i}\sin k_x \sigma_z \nonumber\\
    &+\textrm{i}(\gamma+\cos k_y)\sigma_y+\textrm{i}\sin k_y \sigma_x,
    \label{eq:HOSE_TRS}
\end{align}
where the $-\textrm{i}\gamma \sigma_0$, present in Eq.~\eqref{eq:HOSE}, has been removed, with no consequence for the Hamiltonian's topological phase.
$\textrm{i}h_{\text{HOSE}}(\textbf{k})$ also possesses $C_2$ symmetry with $\hat{r}_2=\sigma_y$. Multiplying $h_\mathrm{HOSE}$ by $\textrm{i}$ rotates its energy spectrum in the complex energy plane by $90$ degrees counter-clockwise. Therefore, $\textrm{i} h_{\text{HOSE}}(\textbf{k})$ has the same topological properties as $h_{\text{HOSE}}(\textbf{k})$. The Hamiltonian $\textrm{i} h_{\text{HOSE}}(\textbf{k})$ is equivalent to Hamiltonian~\eqref{eq:DeformationImGap} at the end of deformation process (i.e. at $g=t=0$) by identifying $\gamma$ in Eq.~\eqref{eq:HOSE_TRS} with $m$ in Eq.~\eqref{eq:DeformationImGap}.

 \begin{table}[h!]
\begin{tabular}{lrrrrr}
  \toprule
  Phase &Band & C & $[X_1^{(2)}]$ & $[Y_1^{(2)}]$ & $[M_1^{(2)}]$ \\
  \midrule
\multirow{2}{*}{HOSE} 
& $\Tilde{1}$ &0 & 0     & -1       & -1  \\
&$-\Tilde{1}$ &0 & 0     & 1        & 1   \\
\midrule
\multirow{2}{*}{Trivial}
&$\Tilde{1}$  &0 & 0     & 0        & 0   \\
&$-\Tilde{1}$ &0 & 0     & 0        & 0   \\
\bottomrule
\end{tabular}
\caption{$\chi^{(2)}$ index for Hamiltonian \eqref{eq:HOSE_TRS} in the HOSE phase ($0<\gamma<1$) and trivial phase ($\gamma>1$).}
\label{tab:chi2 HOSE}
\end{table}

The spectrum of $\textrm{i} h_{\text{HOSE}}(\textbf{k})$ under OBC along $x$ and PBC along $y$ is shown in Fig.~\ref{fig:DeformationImGap}(d) for $\gamma=0.5$. We label the top (bottom) energy band in the complex plane with $\Tilde{1}$ ($-\Tilde{1}$). 

For $0<\gamma<1$, the model is in a HOSE phase. For $\gamma>1$, it is in a trivial phase. The $\chi^{(2)}$ indices for this model are shown in Table \ref{tab:chi2 HOSE} for both phases.

\section{Spectra of HOTK phases with $C_3$ and $C_6$ symmetries}\label{appendix:C3 and C6 spectrum}
\begin{figure*}[t!]
    \centering
    \includegraphics[width=\textwidth]{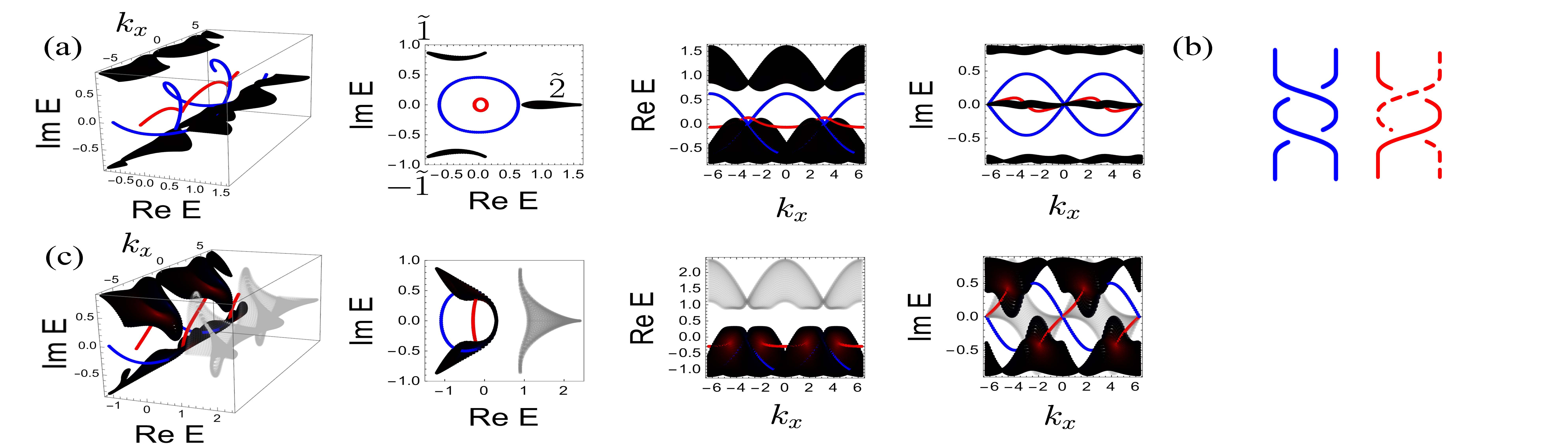}
    \caption{(a)[(c)] Energy spectrum of Hamiltonian \eqref{eq:h3} plotted under PBC along $x$, OBC along $y$ plotted at $t=0.3$ [$t=0.7$] with zigzag edge. (b) Schematic of the braid structure that the edge energy bands trace in the complex energy plane as they traverse the 1D BZ. The dashed line represents the $E=(0,0)$ point around which the red edge state winds. 
    In (c), band $\Tilde{2}$ is plotted in gray to facilitate the visibility of edge states. Blue and red colors correspond to opposite edge states.}
    \label{fig:NH_Kagome}
\end{figure*}

\begin{figure*}[t!]
    \centering
    \includegraphics[width=\textwidth]{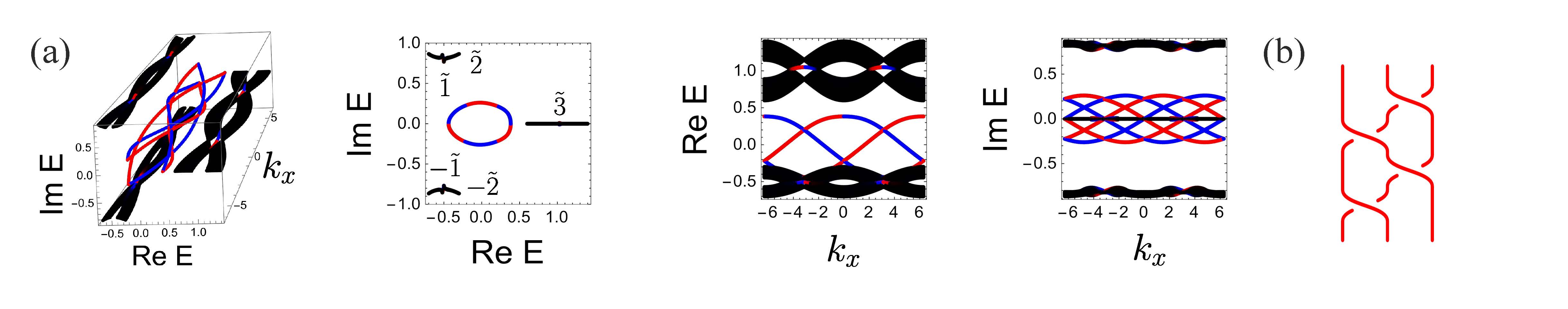}
    \caption{ (a) Energy spectrum of Hamiltonian \eqref{eq:h6} with PBC along $x$ and OBC along $y$ (with zigzag edges), for $t=0.2$. (b) Schematic of the braid structure that edge energy bands trace in the complex energy plane as they traverse the 1D BZ (shown only for one edge, the other edge has the same braid structure).}
    \label{fig:NH hexagon 2}
\end{figure*}

This appendix contains Figs.~\ref{fig:NH_Kagome} and \ref{fig:NH hexagon 2} with plots of the energy bands of Hamiltonians~\eqref{eq:h3}~and~\eqref{eq:h6} that complement the information shown in the main text. In both figures, energy bands are plotted under PBC along $x$, OBC along $y$.

Fig.~\ref{fig:NH_Kagome} (a)[(c)] shows the energy spectra of Hamiltonian~\eqref{eq:h3} at the values $t=0.3$ [$t=0.7$]. 
In (a), the Hamiltonian~\eqref{eq:h3} is in the HOTK phase. The knot formed by the edge states in this phase is shown in (b). In (c), the Hamiltonian~\eqref{eq:h3} is in a imaginary-line-gap complex Chern insulator phase, where chiral edge states close the imaginary line gap between bands $\pm\Tilde{1}$. These two phases are separated by a phase transition around $t=0.57$ that closes the line gap between bands $\Tilde{1}$, $\Tilde{2}$ and $-\Tilde{1}$, $\Tilde{2}$.

Fig.~\ref{fig:NH hexagon 2} (a) shows the energy spectra of Hamiltonian~\eqref{eq:h6} at the value $t=0.2$, where Hamiltonian~\eqref{eq:h6} is in the HOTK phase. The knot formed by the edge states is shown in (b). Hamiltonian~\eqref{eq:h6} does not support the complex Chern insulator phase at any value of $t$. 

\begin{figure*}[t!]
    \centering
    \includegraphics[width=\textwidth]{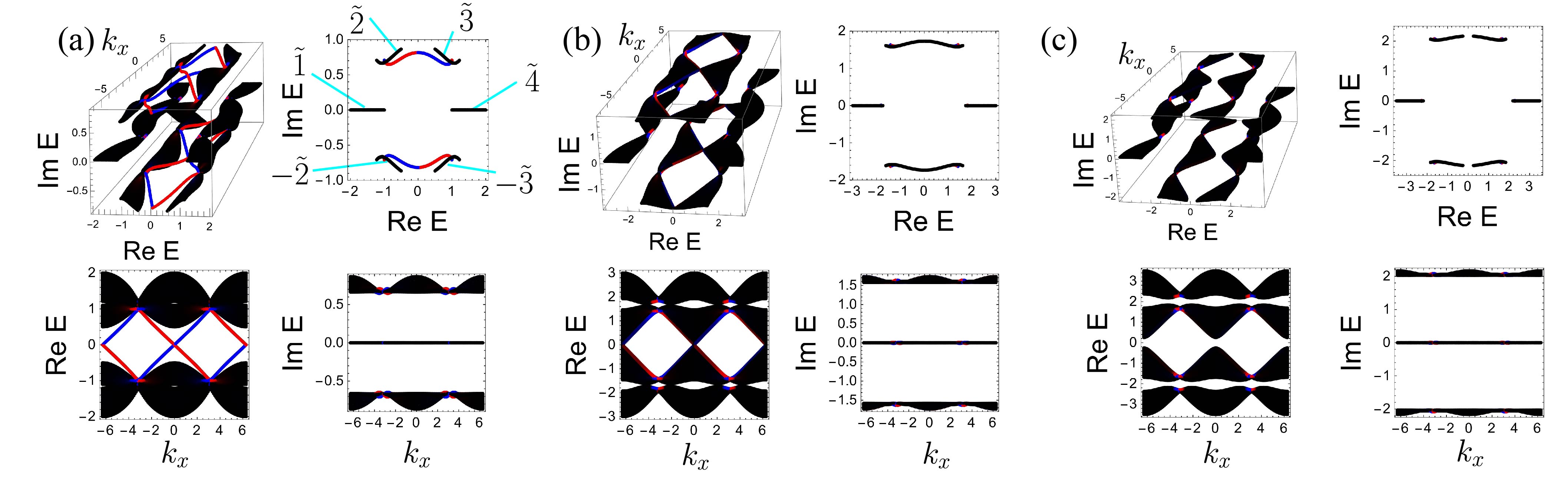}
    \caption{Energy spectrum under OBC along $y$, PBC along $x$ (with zigzag edges) of the NH breathing honeycomb lattice with Hamiltonian \eqref{eq:NH Breathing HoneyComb}. (a) real-line-gap complex Chern insulator phase at $t=1$; (b) phase transition closing the real line gap at $t=2$, (c) trivial phase at $t=2.5$.}
    \label{fig:NH Breathing honeycomb}
\end{figure*}

\pagebreak

\section{A $C_6$-symmetric real-line-gap Chern insulator}\label{Sec:NH Breathing Honeycomb}

In this appendix, we provide a $C_6$ symmetric lattice Hamiltonian that realizes a real-line-gap complex Chern insulator [a minimal, four-band model for this phase is shown in Eq.~\eqref{eq:ComplexChernInsulatorReGap}]. 
Consider the NH breathing honeycomb lattice of Fig.~\ref{fig:Lattices}(d). It has Bloch Hamiltonian
\begin{equation}
    h_{\text{BH}}({\bf k})=\left(
\begin{array}{cccccc}
 0 & 0 & 0 & 0 & t & e^{i \textbf{k}\cdot \textbf{a}_1} \\
 0 & 0 & 0 & e^{-i \textbf{k}\cdot\textbf{a}_2} & 0 & t \\
 0 & 0 & 0 & t & e^{-i \textbf{k}\cdot\textbf{a}_3} & 0 \\
 t & e^{i \textbf{k}\cdot\textbf{a}_2} & 0 & 0 & 0 & 0 \\
 0 & t & e^{i \textbf{k}\cdot\textbf{a}_3} & 0 & 0 & 0 \\
 e^{-i\textbf{k}\cdot \textbf{a}_1} & 0 & t & 0 & 0 & 0 \\
\end{array}
\right).
\label{eq:NH Breathing HoneyComb}
\end{equation}

Hamiltonian~\eqref{eq:NH Breathing HoneyComb} obeys TRS~\eqref{eq:regular TRS} with $\mathcal{T}=\mathcal{K}$ and $C_6$ symmetry with a rotation operator $\hat{r}_6$ that is represented by the matrix that permutes the sites within the unit cells in the lattice in Fig.~\ref{fig:Lattices}(d) upon rotation by $2\pi/6$ about the center of the unit cell.

The energy bands for this Hamiltonian are shown in Fig.~\ref{fig:NH Breathing honeycomb}. For $0<t<2$, the Hamiltonian~\eqref{eq:NH Breathing HoneyComb} is in the complex Chern insulator phase, and its energy bands across the real line gap exhibit nontrivial $\chi^{(6)}$ indices that come in opposite pairs (Table \ref{Tab:2}). As a result, topological edge states cross the line gaps that separate the energy bands with opposite $\chi^{(6)}$ indices [Fig.~\ref{fig:NH Breathing honeycomb}(a)]. At $t=2$, there is a transition point, where the real line gap closes [Fig.~\ref{fig:NH Breathing honeycomb}(b)]. For $t>2$, the system enters the trivial phase, with $\chi^{(6)}={\bf 0}$ for all bands and no edge states between them [Fig.~\ref{fig:NH Breathing honeycomb}(c)]. 

\begin{table}
\begin{tabular}{lrrrrr}
\toprule
    Phase & Band & $C$ & $[M_1^{(2)}]$  & $[K_1^{(3)}]$ &  $[K_2^{(3)}]$ \\
    \midrule
    \multirow{6}{*}{complex Chern} &
      $\Tilde{2}$   &2  & 1  & 0 &  1\\
    & $\Tilde{3}$   &-2 & -1 & 0 &  -1\\
    & $\Tilde{1}$   &0  & 0  & 0 &  0\\
    & $\Tilde{4}$   &0  & 0  & 0 &  0\\
    & $-\Tilde{2}$  &-2 & 1  & 0 &  -1\\
    & $-\Tilde{3}$  &2  & -1 & 0 &  1\\
    \bottomrule
\end{tabular}
\caption{$\chi^{(6)}$ indices for Hamiltonian~\eqref{eq:NH Breathing HoneyComb} in the real-line-gap complex Chern insulator phase, for $t<2$. The bands are labeled as indicated in Fig.~\ref{fig:NH Breathing honeycomb}(a). For $t>2$, all the bands are trivial, with $\chi^{(6)}={\bf 0}$.}
\label{Tab:2}
\end{table}

\newpage

\bibliography{MyLibrary.bib}

\begin{thebibliography}{79}
\bibitem [{foo()}]{footnote1}%
  \BibitemOpen
  \href@noop {} {}\bibinfo {note} {An extra term $-\textrm{i}\gamma\sigma_0$
  will appear in this derivation, but this term only shifts the spectrum of the
  Hamiltonian~\eqref{eq:HOSE} in the complex energy plane, with no consequences
  for its topological phase, and can therefore be discarded.}\BibitemShut
  {Stop}%
\end{thebibliography}
\end{document}